%% file: submitted.tex
\documentclass[preprint,12pt]{elsarticle}

\biboptions{longnamesfirst}

\usepackage{amsmath}
\usepackage{amssymb}
\usepackage{amsfonts}

\let\oldsqrt\sqrt
\def\sqrt{\mathpalette\DHLhksqrt}
\def\DHLhksqrt#1#2{%
\setbox0=\hbox{$#1\oldsqrt{#2\,}$}\dimen0=\ht0
\advance\dimen0-0.2\ht0
\setbox2=\hbox{\vrule height\ht0 depth -\dimen0}%
{\box0\lower0.4pt\box2}}

\newtheorem{theorem}{Theorem}

\newtheorem{definition}{Definition}

\journal{arXiv.org}

\begin{document}
\begin{frontmatter}
\title{Maximum A Posteriori Covariance
  Estimation\\ Using a Power Inverse Wishart Prior}
\author[sfn]{S{\o}ren Feodor Nielsen\corref{cor1}} 
\ead{sfn.mes@cbs.dk} 
\cortext[cor1]{Corresponding author}
\address[sfn]{Copenhagen Business School, Solbjerg Plads 3, DK-2000 Frederiksberg, Denmark}
\author[jsp]{Jon Sporring} 
\address[jsp]{eScience center, Department of Computer Science, University of Copenhagen, Universitetsparken 1, DK-2100 Copenhagen, Denmark}

\begin{abstract}
  The estimation of the covariance matrix is an initial step in many
  multivariate statistical methods such as principal components
  analysis and factor analysis, but in many practical applications the
  dimensionality of the sample space is large compared to the number
  of samples, and the usual maximum likelihood estimate is poor.
  Typically, improvements are obtained by modelling or regularization.
  From a practical point of view, these methods are often
  computationally heavy and rely on approximations.  As a fast
  substitute, we propose an easily calculable maximum a posteriori
  (MAP) estimator based on a new class of prior distributions
  generalizing the inverse Wishart prior, discuss its properties, and
  demonstrate the estimator on simulated and real data.
\end{abstract}

\end{frontmatter}

\section{Introduction}

The problem of estimating a large covariance matrix with limited
amounts of data occurs in many different applications of statistics
such as image analysis, functional data analysis, quantitative
finance, analysis of microarray data etc. We became interested in this
problem through the study of shape variations in medical applications,
e.g.\ X-ray images of human vertebra \cite{crimi10}. To study the
shape variation in such data, images are annotated by a medical
expert, and in the case of the vertebra 50 anatomically meaningful
points were set on each 2 dimensional X-ray image, such that each
shape is represented by a 100 dimensional vector.  For such a
high-dimensional space, the standard ML covariance matrix estimate
requires in the order of 1000 annotated images to be of reasonable
accuracy.  Unfortunately, this is rarely available, since the
annotation task is laboursome and medical experts are a limiting
resource.  Therefore, we have been looking into improved estimates for
small samples of high dimension.

In this paper we propose a maximum a posteriori (MAP) estimator for
the unknown covariance matrix based on a new class of prior
distributions, which we call the power inverse Wishart distributions.
We introduce the distributions in section \ref{sec:powerInverseWishart}
and derive the MAP estimator in section \ref{sec:MAP}.  We compare its
properties with those of the usual inverse Wishart MAP estimator in
section \ref{sec:compare}, derive some asymptotic results in section
\ref{sec:asymptotics}, and demonstrate its applicability on simulated
(section \ref{sec:simulations}) as well as on real data (section
\ref{sec:notch.shape}).

\section{The Power Inverse Wishart Distribution}\label{sec:powerInverseWishart}

We start by defining a class of distributions on the set of positive
definite $p\times p$-matrices. This class generalizes the
well-known inverse Wishart distribution and, as we will argue in the
following section, leads to tractable MAP estimators of an unknown
covariance matrix of a multivariate normal distribution.

\begin{definition}
  The power inverse Wishart distribution with parameters
  $(\boldsymbol\Psi,{m},q)$, where $\boldsymbol\Psi$ is a positive
  definite $p\times p$-matrix, ${m}\ge p$, and
  $q\in\{1,2,\ldots\}$, is the distribution on the set of positive
  definite $p\times p$-matrices with density given by
  \begin{equation}
    \label{eq:qiw.density}
    \mathcal{W}^{-q}\hspace{-2pt}\left({\boldsymbol B|\boldsymbol\Psi,{m}}\right)
    =\frac 1{c_{{m},q}}\exp\left(-\frac
      12\operatorname{tr}\left({\left(\boldsymbol{\Psi}^{-1/2}\boldsymbol{B}\boldsymbol{\Psi}^{-1/2}\right)^{-q}}\right)\right)
    \frac{\left|{\boldsymbol\Psi}\right|^{qm/2}}{\left|{\boldsymbol B}\right|^{q{m}/2+p/2+1/2}},
  \end{equation}
  where $c_{{m},q}$ is a normalization constant given by 
  \begin{equation}\label{eq:normalization.integral}
    c_{{m},q}=\int\exp\left(-\frac 12\operatorname{tr}\left({\boldsymbol{B}^{-q}}\right)\right)\left|{\boldsymbol
      B}\right|^{-q{m}/2-p/2-1/2}d\boldsymbol B,
  \end{equation}
  where the integral is over the set of positive definite
  $p\times p$-matrices.
\end{definition}

The distribution is well-defined, when the integral in
\eqref{eq:normalization.integral} is finite; we show this in the
following theorem. For $q=1$, the power inverse Wishart distribution is
the well-known inverse Wishart distribution with density
\begin{equation}\label{eq:invers.Wishart.density}
  \mathcal{W}^{-1}\hspace{-2pt}\left({\boldsymbol{B}|\boldsymbol{\Psi},m}\right)
  = \frac{\left|{\boldsymbol{\Psi}}\right|^{{m}/2}
    \exp\left(-\frac{1}{2}\operatorname{tr}\left({\boldsymbol{\Psi}\boldsymbol{B}^{-1}}\right)\right)}
  {2^{{m} p/2}\left|{\boldsymbol{B}}\right|^{({m}+p+1)/2}
    \Gamma_p\left(\frac{{m}}{2}\right)},
\end{equation}
where
$\Gamma_p\left(\frac{{m}}{2}\right)=\pi^{p(p-1)/4}\prod_{i=1}^p{\Gamma\left(\frac{{m}}{2}-\frac{(i-1)}{2}\right)}$
is the multivariate gamma function. For $p=1$ the power inverse
Wishart distribution is the distribution of $Y^{-q}$ where
$Y/\Psi\sim\chi^2_{(m)}$.

\begin{theorem}\label{thm:density}
  The function given in \eqref{eq:qiw.density} is a density on
  the set of positive definite $p\times p$-matrices.
\end{theorem}

As a preliminary for the proof, recall that any positive definite matrix $\boldsymbol C$ has a positive
definite $q$th root given by $\boldsymbol{C}^{1/q}=\boldsymbol{V}_{\!\!\boldsymbol
  C}\boldsymbol{\Delta}^{1/q}\boldsymbol{V}_{\!\!\boldsymbol C}^\top$ where
$\boldsymbol{V}_{\!\!\boldsymbol C}$ is a orthonormal matrix diagonalizing $\boldsymbol C$,
$\boldsymbol{\Delta}=\boldsymbol{V}_{\!\!\boldsymbol{C}}^\top\boldsymbol{C}\boldsymbol{V}_{\!\!\boldsymbol{C}}$
is the diagonal matrix of eigenvalues of $\boldsymbol C$ and
$\boldsymbol{\Delta}^{1/q}$ is the diagonal matrix with the $q$th root of the
eigenvalues of $\boldsymbol C$ in the diagonal (see, e.g.\ \citet[Appendix A]{mardia79}).
\\\\
\noindent\emph{Proof}  It follows from \citet[Theorem~3.7]{deemer51}
that 
\begin{align*}
   \mathcal{W}^{-q}\hspace{-2pt}\left({\boldsymbol B|\boldsymbol I,{m}}\right)= \mathcal{W}^{-q}\hspace{-2pt}\left({\boldsymbol\Psi^{1/2}\boldsymbol B\boldsymbol\Psi^{1/2}|\boldsymbol\Psi,{m}}\right)\cdot\left|{\boldsymbol\Psi^{1/2}}\right|^{p+1}.
\end{align*}
Thus it is sufficient to show that \eqref{eq:qiw.density}
is a density for $\boldsymbol\Psi=\boldsymbol I$.

Let $\boldsymbol{C}$ be an inverse Wishart-distributed matrix with parameters
$\boldsymbol I$ and ${m}\ge p$, and consider the density of the
distribution of the positive definite $q$th root $\boldsymbol
B=\boldsymbol{C}^{1/q}$ of $\boldsymbol C$,
\begin{equation*}
  \mathcal{W}^{-1}\hspace{-2pt}\left({\boldsymbol{B}^q|\boldsymbol{I},m}\right)\cdot\left|{J(\boldsymbol{B}^q,\boldsymbol{B})}\right|,
\end{equation*}
where $J(\boldsymbol{B}^q,\boldsymbol{B})$ is the Jacobian matrix of
the transformation $h(\boldsymbol B)=\boldsymbol{B}^q$ defined on the
set of symmetric matrices. It follows from \citet[p.\ 438 \&
Lemma~4.5(vi)]{magnus} that
\begin{align*}
  \left|{J(\boldsymbol B^q,\boldsymbol{B})}\right|=q^p\left|{\boldsymbol B}\right|^{q-1}
  \prod_{i<j}\frac{\lambda_i^q-\lambda_j^q}{\lambda_i-\lambda_j},
\end{align*}
where $\lambda_1>\lambda_2>\ldots>\lambda_p>0$ are the
eigenvalues of $\boldsymbol B$.The last term may be bounded from below as
follows:
\begin{align*}
  \frac{\lambda_i^q-\lambda_j^q}{\lambda_i-\lambda_j}
  &=\lambda_i^{q-1}\frac{1-(\lambda_j/\lambda_i)^q}{1-\lambda_j/\lambda_i}
  =\lambda_i^{q-1}\sum_{l=0}^{q-1}(\lambda_j/\lambda_i)^l\\
  &\ge \frac 1{\max_{l=0,\ldots,q-1}\binom{q-1}{l}}
  \sum_{l=0}^{q-1}\binom{q-1}{l}\lambda_j^l\lambda_i^{q-1-l}\\
  &=\frac {(\lambda_i+\lambda_j)^{q-1}}{\max_{l=0,\ldots,q-1}\binom{q-1}{l}}
  \ge\frac {2^{(q-1)/2}}{\max_{l=0,\ldots,q-1}\binom{q-1}{l}}\cdot \sqrt{\lambda_i\lambda_j}\,^{q-1}.
\end{align*}
Thus
\begin{align*}
   \left|{J(\boldsymbol{B}^q,\boldsymbol{B})}\right|\ge\emph{const}\cdot\left|{\boldsymbol B}\right|^{q-1+(q-1)(p-1)/2}.
\end{align*}
Hence the density of $\boldsymbol B={\boldsymbol C}^{1/q}$ bounds
\begin{align*}
  \exp\left(-\frac 12\operatorname{tr}\left({\boldsymbol{B}^{-q}}\right)\right)\left|{\boldsymbol B}\right|^{-q/2({m}+p+1)+q-1+(q-1)(p-1)/2}\\=\exp\left(-\frac 12\operatorname{tr}\left({\boldsymbol{B}^{-q}}\right)\right)\left|{\boldsymbol B}\right|^{-q{m}/2-p/2-1/2}
\end{align*}
up to a constant. It follows that \eqref{eq:qiw.density} is
integrable, and therefore it specifies a density.\qed
\\\\
The next result, which describes the standard (i.e. $\boldsymbol\Psi=\boldsymbol I$)
power inverse Wishart distribution, follows directly from
\citet[Theorem 13.3.4]{anderson03}:
\begin{theorem}
  Suppose $B$ is  a power inverse Wishart $(\boldsymbol\Psi,{m},q)$-distributed
  $p\times p$-matrix and let
  $\lambda_1>\lambda_2>\ldots\lambda_p>0$ denotes its eigenvalues
  and $\boldsymbol V\!$ the matrix containing its normalised eigenvectors
  chosen such that the first element of each column is non-negative.
  
  Then $(\lambda_1,\ldots,\lambda_p)$ and $\boldsymbol V \!$ are
  independent, the joint density of  the eigenvalues is
  \begin{equation*}
    g(\lambda_1,\ldots,\lambda_p)=\dfrac{\pi^{p^2/2}}{c_{{m},q}\Gamma_p(p/2)}\cdot\dfrac{\exp\left(-\frac 12\sum_{i=1}^p\lambda_i^{-q}\right)}{\prod_{i=1}^p\lambda_i^{({m}+p+1)/2}}\cdot\prod_{i<j}(\lambda_i-\lambda_j),
  \end{equation*}
  and $\boldsymbol V$ has the conditional Haar invariant distribution (cf
  \citet[Definition~13.3.1]{anderson03}).
\end{theorem}
The theorem says that the eigenvectors of a power inverse Wishart
distributed matrix (including the inverse Wishart distribution) with
$\boldsymbol\Psi=\boldsymbol I$ have the same distribution as the eigenvectors of a
Wishart distributed matrix with the same matrix-parameter. Hence the
distributions differ in how the eigenvalues are distributed. 
\\\\
It follows from \citet[Lemma~4.2.1]{mardia79} that the mode of the
power inverse Wishart distribution is
\begin{equation}
  \label{eq:prior.mode}
  \left(\frac q{q{m}+p+1}\right)^{1/q}\!\!\cdot{\boldsymbol\Psi}.
\end{equation}
To compare the power inverse Wishart distribution to the inverse
Wishart distribution, we look at the ratio
\begin{equation*}
  \frac{\mathcal{W}^{-q}\hspace{-2pt}\left({\boldsymbol B|\boldsymbol I,{m}_q}\right)}{\mathcal{W}^{-1}\hspace{-2pt}\left({\boldsymbol B|\boldsymbol
      I,{m}_1}\right)}=\emph{const}\cdot\prod_{i=1}^p\exp\left(-\frac {1}2\left(\lambda_i^{-q}-\lambda_i^{-1}\right)\right)\lambda_i^{-(q{m}_q-{m}_1)/2}.
\end{equation*}
Here $\lambda_1,\ldots,\lambda_p$ denotes the eigenvalues of
$\boldsymbol B$. We see that as any $\lambda_i\to 0$, this ratio goes to 0.
Thus used as a prior for an unknown positive definite matrix, the
general power inverse Wishart distribution gives smaller credibility
to small eigenvalues, than does the usual inverse Wishart prior, and
this effect gets stronger for larger values of $q$. The behaviour of
the ratio as $\lambda_i\to\infty$ is determined by the parameters
${m}_1$ and ${m}_q$ as well as by $q$: If
$q{m}_q>{m}_1$, then the power inverse Wishart
will penalise large eigenvalues harder, than the inverse Wishart does,
whereas it will be more lenient if $q{m}_q<{m}_1$.
If $q{m}_q={m}_1$, then the ratio will approach a
constant as $\lambda_i\to\infty$. Similar comments can be made in the
case with a general $\boldsymbol\Psi$; in this case the eigenvalues
$\lambda_1,\ldots,\lambda_p$ denotes the eigenvalues of
$\boldsymbol\Psi^{-1/2}\boldsymbol B\boldsymbol\Psi^{-1/2}$. Thus $\boldsymbol\Psi$ is a ``scaling
parameter'' and determines the position of the distribution, whereas
$q$ determines the tail behaviour at the ``lower tail'', and the
product $qm$ determines the upper tail behaviour.

\begin{figure}
  \centering
  \includegraphics[width=\textwidth]{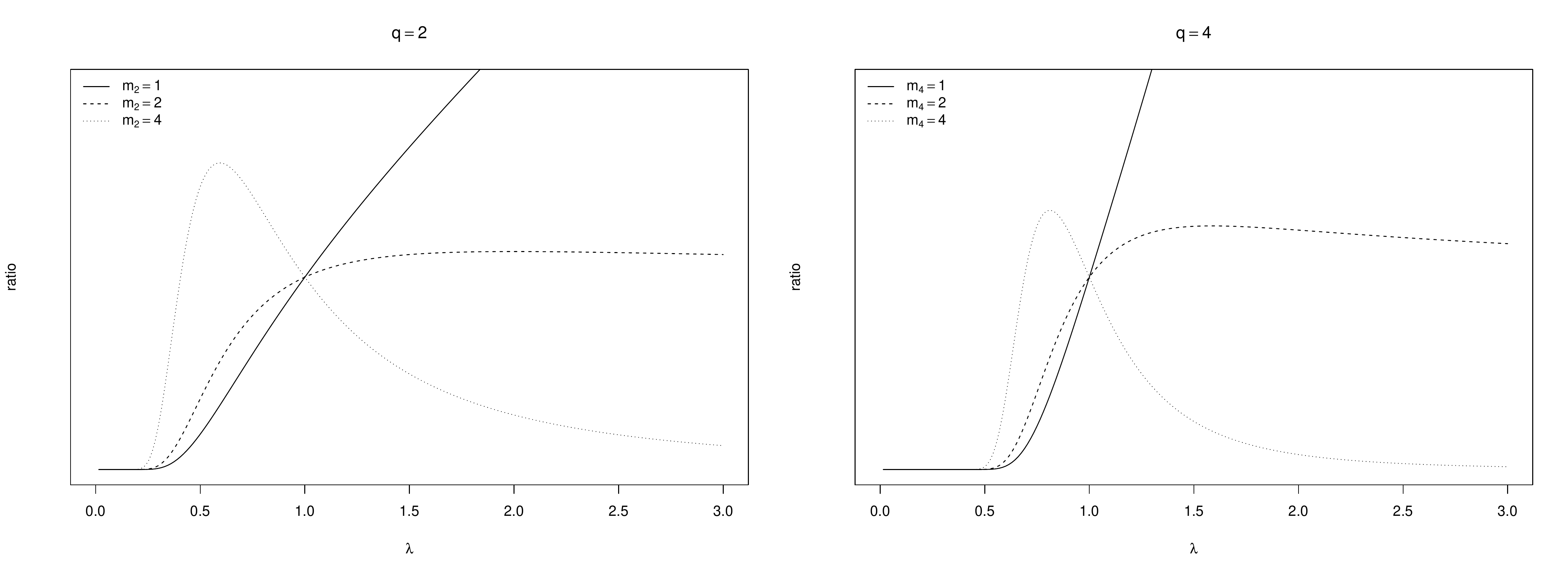}
  \caption{Ratio of the power inverse Wishart density to the inverse
    Wishart density for different values of the parameters.  The left
    hand graph shows ratios for $q=2$ and ${m}_1=4$, the
    right hand graph shows ratios for $q=4$ and ${m}_1=8$.
    In both graphs ratios are given for ${m}_q=1$, $2$ and
    $4$. The ratios have been normalized to take the same value at
    $\lambda=1$.}
  \label{fig:ratios}
\end{figure}

\begin{figure}
  \centering
  \includegraphics[width=\textwidth]{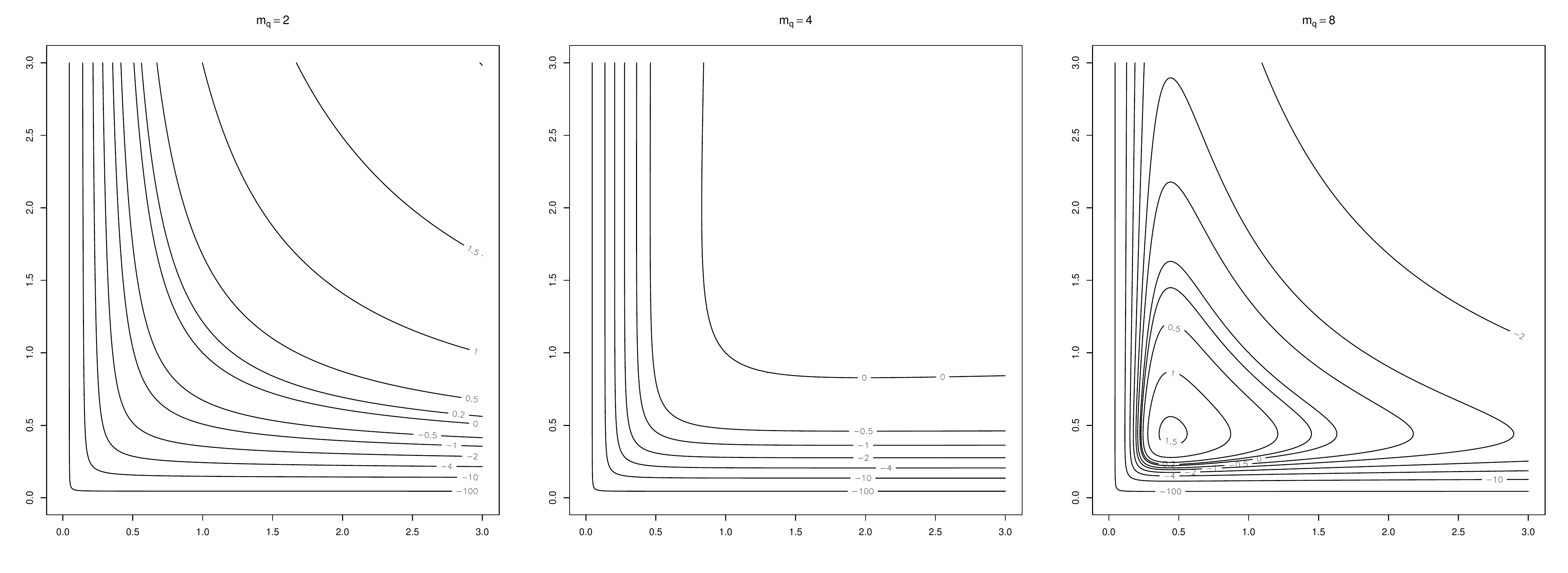}
  \caption{Ratio of the 2-power inverse Wishart density with
    ${m}_q=2$, $4$ and $8$ to the inverse Wishart density
    with ${m}_1=8$. The ratios are normalized
    to take the value one at $(1,1)$. The level curves are drawn
    at $10^{c}$ with the value of $c$ ($-100,-10,
    -4,-2,-1,-0.5,0,0.2,0.5,1,1.5,2$) denoted on the graphs.}
  \label{fig:ratios.levelcurves}
\end{figure}

We illustrate the tail behaviour in figure~\ref{fig:ratios} for
$p=1$ and in figure~\ref{fig:ratios.levelcurves} for $p=2$
by plotting the ratios or the level curves of the ratios of the power
inverse Wishart density to the inverse Wishart density for
selected values of the parameters.

\section{Maximum A Posteriori Estimation}
\label{sec:MAP}
Consider a random sample $\boldsymbol{X}_1,\ldots,\boldsymbol{X}_n\in\mathbb{R}^p$ of $n$
independent and identically normally distributed $p$-dimensional
random vectors, where both the mean vector $\boldsymbol{\mu}$ and the
covariance matrix $\boldsymbol{\Sigma}$ are unknown.  The covariance matrix
$\boldsymbol{\Sigma}$ is symmetric, and we will assume it to be positive
definite.  Put $\boldsymbol{\bar X}=\frac 1n\sum_{i=1}^n\boldsymbol{X}_i$
and let
\begin{equation*}
  \boldsymbol{S}
  =\frac {1}{n}\sum_{i=1}^n(\boldsymbol{X}_i-\boldsymbol{\bar
    X})(\boldsymbol{X}_i-\boldsymbol{\bar X})^\top 
\end{equation*}
denote the empirical covariance matrix.  Then the likelihood function
is given by
\begin{align*}
  L\left(\boldsymbol{\mu},\boldsymbol{\Sigma}|\boldsymbol{X}_1,\ldots,\boldsymbol{X}_n\right)&=\dfrac{\exp\left(-\frac 12
      \sum_{i=1}^n(\boldsymbol{X}_i-\boldsymbol{\mu})^\top\boldsymbol{\Sigma}^{-1}(\boldsymbol{X}_i-\boldsymbol{\mu})\right)}{|\boldsymbol{\Sigma}|^{n/2}}\\
  &={|\boldsymbol{\Sigma}|^{-n/2}}{\exp\left(-\frac n 2
      \operatorname{tr}\left({\boldsymbol{\Sigma}^{-1}\boldsymbol{S}}\right)\right)}\\
  &\qquad\qquad\qquad\cdot\exp\left(-\frac n2(\boldsymbol{\bar X}-\boldsymbol{\mu})^\top\boldsymbol{\Sigma}^{-1}(\boldsymbol{\bar X}-\boldsymbol{\mu})\right).
\end{align*}
Provided that $n>p$, the likelihood function has a unique maximum at
\begin{equation*}
  \label{eq:MLE}
  \hat{\boldsymbol{\mu}}=\bar{\boldsymbol{X}},\qquad\hat{\boldsymbol{\Sigma}}=\boldsymbol{S}.
\end{equation*}

If $n\le p$ the likelihood is unbounded, and in this
case there is no maximum likelihood estimate (MLE). Of course
$\bar{\boldsymbol{X}}$ and $\boldsymbol{S}$ may still be used as estimators, but the
properties of these estimators are typically poor. In many
applications it may also be problematic that $\boldsymbol{S}$ is not positive
definite. This is also the case when using methods such as principal
components analysis or factor analysis. Even if the intention here is
to reduce dimensionality, we would generally not want the reduction to
be based simply on insufficient amounts of data leading to a singular
covariance matrix.  Moreover, if $p$ is much larger than
$n$, then the largest eigenvalue of $\boldsymbol S$ may severely
overestimate the largest eigenvalue of $\boldsymbol\Sigma$ even if
$n$ is large (see section~\ref{sec:asymptotics}). One way of
mending these problems is to put a prior distribution on the unknown
parameters and use maximum a posteriori estimators. A standard choice
of prior for $\boldsymbol{\Sigma}$ is the inverse Wishart distribution with
parameters $(\boldsymbol{\Psi},m)$.  With an improper uniform prior on
$\mathbb{R}^p$ for $\boldsymbol{\mu}$ this leads to MAP estimators given by
\begin{equation*}
  \label{eq:IWMAP}
  \hat{\boldsymbol{\mu}}=\bar{\boldsymbol{X}},\qquad\hat{\boldsymbol{\Sigma}}=\frac 1{n+{m}+p+1}(n\boldsymbol{S}+\boldsymbol{\Psi}).
\end{equation*}
Without prior knowledge, a simple choice for the hyperparameter
$\boldsymbol{\Psi}$ would be $\alpha \boldsymbol{I}$ for some
$\alpha$. This leads to an estimator of $\boldsymbol\Sigma$, which has the same
eigenvectors as the MLE, but where the eigenvalues have been scaled
down by $n/(n+{m}+p+1)$ and shifted
upwards by $\alpha/(n+{m}+p+1)$. Thus, every
eigenvalue of $\boldsymbol{S}$ is regularized in the same way regardless of
its size. In some applications it may be more reasonable to apply
different amounts of regularization depending on the size of the eigenvalue.

Instead of using an inverse Wishart prior for the unknown covariance
matrix, $\boldsymbol{\Sigma}$, we propose to use a power inverse Wishart
distribution as prior.  Keeping the improper uniform prior for
$\boldsymbol{\mu}$, the resulting posterior is given by
\begin{align*}
  &\pi\left(\boldsymbol{\Sigma},\boldsymbol{\mu}|\boldsymbol{X}_1,\ldots,\boldsymbol{X}_n\right)\propto
  L\left(\boldsymbol{\mu},\boldsymbol{\Sigma}|\boldsymbol{X}_1,\ldots,\boldsymbol{X}_n\right)
  \cdot\mathcal{W}^{-q}\hspace{-2pt}\left({\boldsymbol{\Sigma}|\boldsymbol{\Psi},m}\right)\\[.5em]
  &\propto\dfrac{\exp\left(-\frac n 2
      \operatorname{tr}\left({\boldsymbol{\Sigma}^{-1}\boldsymbol{S}}\right)\right)\cdot\exp\left(-\frac n2(\boldsymbol{\bar
        X}-\boldsymbol{\mu})^\top\boldsymbol{\Sigma}^{-1}(\boldsymbol{\bar X}-\boldsymbol{\mu})\right)}{\left|{\boldsymbol{\Sigma}}\right|^{1/2}}\\[.5em]
    &\qquad\qquad\qquad\qquad\qquad\qquad\cdot\dfrac{\exp\left(-\frac 12\operatorname{tr}\left({\left(\boldsymbol{\Psi}^{-1/2}\boldsymbol{\Sigma}\boldsymbol{\Psi}^{-1/2}\right)^{-q}}\right)\right)}{|\boldsymbol\Sigma|^{(n+p+q{m})/2}}\\[.2em]
    &\propto\dfrac{\exp\left(-\frac 12\operatorname{tr}\left({n
        \boldsymbol{\Psi}^{-1/2}\boldsymbol{S}\boldsymbol{\Psi}^{-1/2}\cdot\left(\boldsymbol{\Psi}^{-1/2}\boldsymbol{\Sigma}\boldsymbol{\Psi}^{-1/2}\right)^{-1}+\left(\boldsymbol{\Psi}^{-1/2}\boldsymbol{\Sigma}\boldsymbol{\Psi}^{-1/2}\right)^{-q}}\right)\right)}{\left|{\boldsymbol{\Psi}^{-1/2}\boldsymbol{\Sigma}\boldsymbol{\Psi}^{-1/2}}\right|^{({n+p+q{m}+1})/2}}\\[.2cm]
  &\qquad\qquad\qquad\qquad\qquad\qquad\cdot \exp\left(-\frac n2(\bar{\boldsymbol{X}}-\boldsymbol{\mu})^\top
    \boldsymbol{\Sigma}^{-1}(\bar{\boldsymbol{X}}-\boldsymbol{\mu})\right).
\end{align*}
Maximizing over $\boldsymbol{\mu}$ gives us $\hat{\boldsymbol{\mu}}=\bar{\boldsymbol{X}}$.
In order to maximize over $\boldsymbol{\Sigma}$ we put
$\boldsymbol{\mu}=\bar{\boldsymbol{X}}$, change parametrization to
$\boldsymbol{\Upsilon}=\left(\boldsymbol{\Psi}^{-1/2}\boldsymbol{\Sigma}\boldsymbol{\Psi}^{-1/2}\right)^{-1}$, and take logs to obtain
\begin{equation}
\label{eq:log.post}
  \begin{split}
      \boldsymbol{\Upsilon}\to l(\boldsymbol{\Upsilon})
  &=\log\pi\left(\boldsymbol{\Sigma},\boldsymbol{\mu}|\boldsymbol{X}_1,\ldots,\boldsymbol{X}_n\right)\\
  &=  -\frac 12\operatorname{tr}\left({n \boldsymbol{\Psi}^{-1/2}\boldsymbol{S}\boldsymbol{\Psi}^{-1/2}\cdot\boldsymbol{\Upsilon}
    +\boldsymbol{\Upsilon}^{q}}\right)
  +\frac{n+p+q{m}+1}2\log\left|{\boldsymbol{\Upsilon}}\right|\\
  &\qquad\qquad\qquad+\emph{const}.
  \end{split}
\end{equation}
Differentiating wrt.\ $\boldsymbol{\Upsilon}$ (see e.g.\
\cite[Chapter 9]{magnus}) gives us
\begin{align*}
 dl(\boldsymbol{\Upsilon})&=-\frac 12\operatorname{tr}\left({n
   \boldsymbol{\Psi}^{-1/2}\boldsymbol{S}\boldsymbol{\Psi}^{-1/2}d\boldsymbol{\Upsilon}+
   q\boldsymbol{\Upsilon}^{q-1} d\boldsymbol{\Upsilon}}\right)\\
 &\qquad\qquad\qquad\qquad+\frac{n+p+q{m}+1}2\frac 1{\left|{\boldsymbol{\Upsilon}}\right|}\operatorname{tr}\left({\boldsymbol{\Upsilon}^{-1}\left|{\boldsymbol{\Upsilon}}\right|d\boldsymbol{\Upsilon}}\right)\\
  &=-\frac 12\operatorname{tr}\left({\left(n \boldsymbol{\Psi}^{-1/2}\boldsymbol{S}\boldsymbol{\Psi}^{-1/2}+ q\boldsymbol{\Upsilon}^{q-1}-(n+p+q{m}+1)\boldsymbol{\Upsilon}^{-1}\right)d\boldsymbol{\Upsilon}}\right),
\end{align*}
which is 0, if
\begin{align}
 n \boldsymbol{\Psi}^{-1/2}\boldsymbol{S}\boldsymbol{\Psi}^{-1/2}+ q\boldsymbol{\Upsilon}^{q-1}-({n+p+q{m}+1})\boldsymbol{\Upsilon}^{-1}
  =0.\label{eq:post.eqn1}
\end{align}
Differentiating again leads to
\begin{align*}
  d^2l(\boldsymbol{\Upsilon})=-\frac 12\operatorname{tr}\left({d\boldsymbol{\Upsilon}^\top\left(q(q-1)\boldsymbol{\Upsilon}^{q-2} d\boldsymbol{\Upsilon}
      +({n+p+q{m}+1})\boldsymbol{\Upsilon}^{-2}
      d\boldsymbol{\Upsilon}\right)}\right),
\end{align*}
so that the Hessian is negative definite. Moreover, by replacing
$\boldsymbol{\Upsilon}$ in \eqref{eq:log.post} by $t\boldsymbol{\Upsilon}$ it is
easily shown that for any fixed $\boldsymbol{\Upsilon}$ the function $t\to
l(t\boldsymbol{\Upsilon})$ tends to minus infinity as $t$ tends to 0 or
infinity. Thus we may conclude that $l(\boldsymbol{\Upsilon})$ has a unique
maximizer, which solves \eqref{eq:post.eqn1} or equivalently 
\begin{equation}\label{eq:post.eqn2}
  n\boldsymbol{\Psi}^{-1/2}\boldsymbol{S}\boldsymbol{\Psi}^{-1/2}\cdot\boldsymbol{\Upsilon}+q\boldsymbol{\Upsilon}^q-({n+p+q{m}+1})\boldsymbol I=0.
\end{equation}
By transposing the terms of this equation, we see that any symmetric
solution, $\hat{\boldsymbol{\Upsilon}}$, to this equation will commute with
$\boldsymbol{\Psi}^{-1/2}\boldsymbol{S}\boldsymbol{\Psi}^{-1/2}$. It follows that
$\boldsymbol{\Psi}^{-1/2}\boldsymbol{S}\boldsymbol{\Psi}^{-1/2}$ and $\hat{\boldsymbol\Upsilon}$
are diagonalized by the same orthonormal matrix (see
\cite[1c(iii)]{rao73}), and consequently the $i$th eigenvalue $\lambda_i$ of
$\hat{\boldsymbol\Upsilon}$ satisfies
\begin{equation}\label{eq:qiw.MAP.eigenvalues}
  q\lambda_i^q+n\lambda_i\left(\boldsymbol{\Psi}^{-1/2}\boldsymbol{S}\boldsymbol{\Psi}^{-1/2}\right)\cdot\lambda_i
  -(n+p+q{m}+1) =0,
\end{equation}
where $\lambda_i(\boldsymbol{\Psi}^{-1/2}\boldsymbol{S}\boldsymbol{\Psi}^{-1/2})$ denotes
the $i$th eigenvalue of
$\boldsymbol{\Psi}^{-1/2}\boldsymbol{S}\boldsymbol{\Psi}^{-1/2}$\!. 

\begin{theorem}
  If we impose a power inverse Wishart prior distribution for
  $\boldsymbol{\Sigma}$ with parameters $({m},\boldsymbol{\Psi},q)$ and an
  improper uniform prior of $\boldsymbol{\mu}$, then the maximum a posteriori
  estimator of $\boldsymbol{\Sigma}$ is
  \begin{equation}\label{eq:qiw.MAP}
    \hat{\boldsymbol\Sigma}=\boldsymbol{\Psi}^{1/2}\boldsymbol{V}\hat{\boldsymbol\Delta}^{-1}\boldsymbol{V}^\top\boldsymbol{\Psi}^{1/2},
  \end{equation}
  where $\hat{\boldsymbol\Delta}$ is a diagonal matrix with the unique positive
  solutions to the equations \eqref{eq:qiw.MAP.eigenvalues} in the
  diagonal, and $\boldsymbol V$ is an orthonormal matrix diagonalizing
  $\boldsymbol{\Psi}^{-1/2}\boldsymbol{S}\boldsymbol{\Psi}^{-1/2}$\!.
\end{theorem}

\noindent\emph{Proof} The polynomial in
\eqref{eq:qiw.MAP.eigenvalues},
\begin{equation*}
  \lambda\to q\lambda^q+n\lambda_i(\boldsymbol\Psi^{-1/2}\boldsymbol S\boldsymbol\Psi^{-1/2})\cdot\lambda
  -(n+p+q{m}+1),
\end{equation*}
is negative for $\lambda=0$ and goes to
infinity as $\lambda\to\infty$. Furthermore, it is strictly increasing
for $\lambda>0$ so that \eqref{eq:qiw.MAP.eigenvalues} has exactly one
positive solution. Hence $\hat{\boldsymbol\Delta}$ is well-defined. Moreover
$\hat{\boldsymbol\Upsilon}=\boldsymbol V^\top\hat{\boldsymbol\Delta}\boldsymbol V$ clearly solves
\eqref{eq:post.eqn2}. It follows that
\begin{align}
  \hat{\boldsymbol\Sigma}=\boldsymbol\Psi^{1/2}\hat{\boldsymbol\Upsilon}^{-1}\boldsymbol\Psi^{1/2}
  =\boldsymbol{\Psi}^{1/2}\boldsymbol{V}\hat{\boldsymbol\Delta}^{-1}\boldsymbol{V}^\top\boldsymbol{\Psi}^{1/2}.\tag*{\qed}
\end{align}
\\
The positive solution of \eqref{eq:qiw.MAP.eigenvalues} is easily
found numerically; we know that it is unique, and by Cauchy's bound
\cite{cauchy} it is bounded by
\begin{align*}
  1+\max\left(n\lambda_i(\boldsymbol\Psi^{-1/2}\boldsymbol S\boldsymbol\Psi^{-1/2}),{n+p+q{m}+1}\right)/q. 
\end{align*}
Hence, we may solve
\eqref{eq:qiw.MAP.eigenvalues} by a numerical method such as bisection.
In the case $q=2$, the eigenvalue equations \eqref{eq:qiw.MAP.eigenvalues} have
closed form solutions
\begin{equation*}\label{eq:qiw.eigenvalues}
  \begin{split}
      \lambda_i^{-1}=\frac n{2({n+p+2{m}+1})}
  \Bigg(\lambda_i(&\boldsymbol\Psi^{-1/2}\boldsymbol S\boldsymbol\Psi^{-1/2})\\
    &+\sqrt{\lambda_i(\boldsymbol\Psi^{-1/2}\boldsymbol S\boldsymbol\Psi^{-1/2})^2+
      8\frac{{n+p+2{m}+1}}{n^2}}\,\Bigg).
  \end{split}
\end{equation*}
It follows that when $q=2$, then
\begin{multline*}
  \label{eq:MAP.estimator}
  \hat{\boldsymbol{\Sigma}}=\frac
  n{2({n+p+2{m}+1})}\left(\boldsymbol{S}+\boldsymbol\Psi^{1/2}\left(\left(\boldsymbol\Psi^{-1/2}\boldsymbol{S}\boldsymbol\Psi^{-1/2}\right)^2\right.\right.\\
  \left.\left.+8\frac{{n+p+2{m}+1}}{n^2}\boldsymbol{I}\right)^{1/2}\,\boldsymbol\Psi^{1/2}\right),
\end{multline*}
which further simplifies to
\begin{equation}\label{eq:SqInvW.MAP}
  \hat{\boldsymbol{\Sigma}}=\frac n{2({n+p+2{m}+1})}\left(\boldsymbol{S}+\left(\boldsymbol{S}^2+8\alpha^2\frac{{n+p+2{m}+1}}{n^2}\boldsymbol{I}\right)^{1/2}\,\right),
\end{equation}
when $\boldsymbol\Psi=\alpha\boldsymbol I$.

\section{Regularization: Floor and shrinkage}\label{sec:compare}

In the previous section we derived the power inverse Wishart MAP,
which includes the usual inverse Wishart MAP as a special case.  In
this section we will discuss and compare how the MAP estimators
regularize the MLE. We will focus mainly on the case, where $\boldsymbol\Psi$ is
a diagonal matrix, as this allows us to give some concrete
expressions, but we will also comment on results for more general
choices of $\boldsymbol\Psi$. 

When $\boldsymbol{\Psi}=\alpha\boldsymbol{I}$ we may write
\begin{equation}\label{eq:special.qiw.MAP}
  \hat{\boldsymbol\Sigma}=\boldsymbol{V}\hat{\boldsymbol\Delta}\boldsymbol{V}^\top,
\end{equation}
where the orthonormal matrix $\boldsymbol V$ diagonalizes $\boldsymbol S$, and
$\hat{\boldsymbol\Delta}$ is the diagonal matrix with diagonal elements given by
the positive solutions to the equations
\begin{equation}
  \label{eq:eigenvals}
  (n+p+q{m}+1)\lambda_i^q-n\lambda_i\left(\boldsymbol{S}\right)\lambda_i^{q-1}
  -q\alpha^q =0.
\end{equation}
In this case, the MLE and the various MAP estimators all share the
same eigenspaces, i.e.\ they are diagonalised by the same orthonormal
matrix $\boldsymbol{V}$\!. The eigenvalues of the MAP estimators are the
diagonal elements of $\hat{\boldsymbol\Delta}$ from
\eqref{eq:special.qiw.MAP}, i.e.\ the solutions to the equations
\eqref{eq:eigenvals}. Thus, the MAP estimators regularizes the
eigenvalues of $\boldsymbol{S}$, but leave the eigenvectors unchanged.  Hence
their difference is, how the eigenvalues are regularized.

If $\lambda$ is an eigenvalue of $\boldsymbol{S}$, then the
corresponding eigenvalue for the inverse Wishart MAP estimator
\eqref{eq:IWMAP} is
\begin{equation}\label{eq:IW.reg.fct}
  \frac 1{n+{m}+p+1}(n\lambda+\alpha),
\end{equation}
and for the 2-power inverse Wishart MAP \eqref{eq:SqInvW.MAP} we get
\begin{equation}\label{eq:SIW.reg.fct}
  \frac n{2({n+p+2{m}+1})}
  \left(\lambda+\sqrt{\lambda^2+
      8\frac{{n+p+2{m}+1}}{n^2}\alpha^2}\,\right).
\end{equation}
Hence, both MAP estimators regularize the MLE by imposing a lower
limit for the eigenvalues, which we denote the {\em floor}, and
shrinking large eigenvalues by multiplying with a factor smaller than
1.  In other words, both MAP estimators increase small eigenvalues and
decrease large eigenvalues as compared to the MLE.  We define the {\em
  shrinkage} as the limit of the regularized eigenvalue divided by the
corresponding unregularized eigenvalue as the latter tends to
infinity. Thus, the shrinkage is the (asymptotic) scaling of large
eigenvalues performed by the MAP estimator, whereas the floor is the
lower limit for small eigenvalues imposed by the MAP estimator. The
floor and the shrinkage factor both improve the estimation: The floor
serves to make the estimator positive definite, whereas shrinking is
beneficial for the estimation of the largest eigenvalues, as these
tend to be overestimated, when $p$ is not negligible compared to
$n$ (see also the following section).

For the inverse Wishart MAP, the floor and the shrinkage are
\begin{equation*}
  \frac \alpha{n+{m}+p+1}\quad\text{and}
  \quad\frac n{n+{m}+p+1}
\end{equation*}
respectively, whereas for the 2-power inverse Wishart the
floor and shrinkage are
\begin{equation*}
\frac{\sqrt{2}\alpha}{\sqrt{{n+p+2{m}+1}}}
  \quad\text{and}\quad\frac n{{n+p+2{m}+1}}
\end{equation*}
respectively.  For general $q$ the floor and shrinkage are
\begin{equation}\label{eq:general.floor.shrink}
  \alpha\left(\frac{ q}{n+p+q{m}+1}\right)^{1/q}
  \quad\text{and}\quad
  \frac n{n+p+q{m}+1}
\end{equation}
respectively. The floor follows directly from
\eqref{eq:eigenvals}, which also shows that
\begin{equation}\label{eq:qiw.eigenval.lower.bound}
  \lambda_i\ge\frac{n}{n+p+q{m}+1}\lambda_i(\boldsymbol S).
\end{equation}
Combining this with Cauchy's bound \cite{cauchy}
\begin{equation}\label{eq:qiw.eigenval.upper.bound}
  \lambda_{i}\le 1+\max\left(\alpha^q q,n\lambda_{i}(\boldsymbol
    S)\right)/(n+p+q{m}+1), 
\end{equation}
we obtain the shrinkage given in \eqref{eq:general.floor.shrink}
above.

The inverse Wishart MAP regularizes the eigenvalues by applying a
linear function to the eigenvalues of $\boldsymbol{S}$; the power inverse
Wishart MAP returns a strictly increasing and strictly convex function
of the eigenvalues of $\boldsymbol{S}$. For $q=2$ this follows directly from
the expression \eqref{eq:SIW.reg.fct}. For general $q$, the Implicit
Function Theorem gives us
\begin{equation}\label{eq:implicit.diff}
  \frac{d\lambda_i}{d\lambda_i(\boldsymbol S)}=\frac n{q(n+p+q{m}+1)-n(q-1)\lambda_i(\boldsymbol S)/\lambda_i},
\end{equation}
which is positive by \eqref{eq:qiw.eigenval.lower.bound}, so that the function is increasing. Differentiating
again we obtain
\begin{equation*}
   \frac{d^2\lambda_i}{d\lambda_i(\boldsymbol
     S)^2}=\frac{n^2(q-1)}{(q(n+p+q{m}+1)-n(q-1)\lambda_i(\boldsymbol S)/\lambda_i)^2}\frac 1{\lambda_i}\left(1-\frac{\lambda_i(\boldsymbol S)}{\lambda_i}\cdot\frac{d\lambda_i}{d\lambda_i(\boldsymbol S)}\right),
\end{equation*}
which is positive, proving convexity. The convex regularization imposed by
the power inverse Wishart prior has the effect that the difference
between small eigenvalues after regularization is smaller than those
between large eigenvalues. Thus the power inverse Wishart MAP
regularizes eigenvalues differently depending on their sizes.

We also note that with the same floor and shrinkage, the eigenvalues
of a power inverse Wishart MAP will always be smaller than the
eigenvalues of the inverse Wishart MAP.  Moreover, as the value of the
derivative \eqref{eq:implicit.diff} at zero is a decreasing function
of $q$, the eigenvalue of a power inverse Wishart MAP corresponding to
any specific eigenvalue of $\boldsymbol S$ is decreasing as a function of the
power $q$, when the floor and shrinkage are unchanged.
\\\\
It is difficult to extend these results to the general case, where
$\boldsymbol\Psi$ is not of the form $\alpha\boldsymbol I$, in a useful way. Clearly
the results may be extended to results concerning the MAP estimator of
$\boldsymbol\Psi^{-1/2}\boldsymbol\Sigma\boldsymbol\Psi^{-1/2}$ by replacing $\boldsymbol S$ with
$\boldsymbol\Psi^{-1/2}\boldsymbol S\boldsymbol\Psi^{-1/2}$, $\boldsymbol\Sigma$ with
$\boldsymbol\Psi^{-1/2}\boldsymbol\Sigma\boldsymbol\Psi^{-1/2}$ and putting $\alpha=1$.
From this we see that the $i$th diagonal element of
$\hat{\boldsymbol\Delta}^{-1}$ is larger than $\beta$ and smaller than
$\beta+\gamma\lambda_i(\boldsymbol\Psi^{-1/2}\boldsymbol S\boldsymbol\Psi^{-1/2})$, where
$\beta$ and $\gamma$ are the floor and shrinkage respectively from
\eqref{eq:general.floor.shrink} with $\alpha=1$. Thus in the usual
ordering of positive semi-definite matrices we have
\begin{align*}
  \beta\boldsymbol I\le\hat{\boldsymbol\Delta}^{-1}\le\beta\boldsymbol I+\gamma\boldsymbol\Delta,
\end{align*}
where ${\boldsymbol\Delta}$ is the diagonal matrix with the eigenvalues of
$\boldsymbol\Psi^{-1/2}\boldsymbol S\boldsymbol\Psi^{-1/2}$ in the diagonal. From this we
obtain
\begin{equation}\label{eq:piw.bounds}
  \beta\boldsymbol\Psi\le\hat{\boldsymbol\Sigma}\le \beta\boldsymbol\Psi+\gamma\boldsymbol S.
\end{equation}
Thus in the general case, we may talk of a ``matrix floor'',
$\beta\boldsymbol\Psi$, and also here there is a shrinkage effect, but the
actual shrinkage may be smaller than the factor $\gamma$.

The inequalities in \eqref{eq:piw.bounds} has two trivial consequences that
may be worth pointing out. The first is that similar inequalities hold
for the diagonal elements of the matrices, i.e.\ for the estimated
variances. The second consequence is that the MAP estimator has
moments of all orders. 

\section{Asymptotic results}\label{sec:asymptotics}
In a standard asymptotic set-up  with $\boldsymbol\Psi$,
${m}$, and $p$ fixed as $n$ increases, the
asymptotic behaviour of a power inverse Wishart MAP is the same as the
asymptotic behaviour of the MLE.

\begin{theorem}
  Suppose that $\boldsymbol\Psi$, ${m}$, and $p$ are fixed as
  $n$ increases. Then the power inverse Wishart MAP
  $\hat{\boldsymbol\Sigma}$ and its eigenvalues have the same asymptotic
  distributions as the MLE $\boldsymbol S$.
\end{theorem}

\noindent\emph{Proof} First consider the case where
$\boldsymbol\Psi=\alpha\boldsymbol I$. As the eigenvalues of $\hat{\boldsymbol\Sigma}$ are
bounded in probability by \eqref{eq:qiw.eigenval.upper.bound}, it
follows that $\hat{\boldsymbol\Sigma}^{q-1}$ is bounded in probability.
Hence, re-writing \eqref{eq:post.eqn2} as
\begin{equation*}
  \left(\boldsymbol
    S-\frac{n+p+q{m}+1}n\hat{\boldsymbol\Sigma}\right)\hat{\boldsymbol\Sigma}^{q-1}=\frac{\alpha^q q}n\boldsymbol I,
\end{equation*}
it follows that $\hat{\boldsymbol\Sigma}=\boldsymbol S+o_P(1/\sqrt n\,)$ and the
result follows.

With a general fixed $\boldsymbol\Psi$, it follows that
$\boldsymbol\Psi^{-1/2}\hat{\boldsymbol\Sigma}\boldsymbol\Psi^{-1/2}=\boldsymbol\Psi^{-1/2}\boldsymbol
S\boldsymbol\Psi^{-1/2}$ $+o_P(1/\sqrt n\,)$, implying that also in this case
the MAP estimator and the MLE have the same asymptotic distribution.

The results concerning the eigenvalues follow by continuous mapping.\qed
\\\\
The densities of the limiting distributions in the case where
$\boldsymbol\Sigma=\boldsymbol I$ are given in \citet[Theorem 13.3.5]{anderson03}. 
\\\\
As indicated in the introduction, our main interest is in estimating the
covariance matrix in situations, where $p$ is large compared to
$n$.  Assuming that the components of $\boldsymbol{X}_i$ are iid
standard normal, and that both $n$ and $p$ increase such
that $n/p\to\gamma\in[0;\infty]$, it is known
that
\begin{align*}
  \frac{\lambda_{\max}(\boldsymbol S)-\mu_{n,p}}{\sigma_{n,p}},
\end{align*}
where $\lambda_{\max}(\boldsymbol S)$ denotes the largest eigenvalue of
$\boldsymbol{S}$, and $\mu_{n,p}$ and
$\sigma_{n,p}$ are given by
\begin{equation}\label{eq:asymp.pars}
  \begin{split}
      \mu_{n,p}&=\left(1+\sqrt{\frac{p}{n}\,}\,\right)^2,\\
  \sigma_{n,p}&=\frac{\sqrt{n}+\sqrt{p}}n\,
  \left(\frac 1{\sqrt{n}}+\frac 1{\sqrt{p}}\right)^{1/3},
\end{split}
\end{equation}
converges in distribution to a Tracy-Widom distribution
\cite{johnstone,karoui}.  For the MAP estimators we show the following
result:

\begin{theorem}\label{thm:tracy-widom}
  Suppose that $\boldsymbol{X}_1,\ldots,\boldsymbol{X}_n$ are independent, standard normally 
  distributed random variables.  Let
  $\lambda_{\max}^{(q)}$ denote the largest eigenvalue of the MAP
  estimator of $\boldsymbol\Sigma$ based on an power inverse Wishart prior
  with parameters $(\alpha\boldsymbol I,{m},q)$. Then
  with
  $\mu_{n,p}$ and $\sigma_{n,p}$ as in
  \eqref{eq:asymp.pars},
  \begin{equation*}
    \frac{\lambda_{\max}^{(q)}-\frac
      n{n+p+q{m}+1}\,\mu_{n,p}}{\frac
      n{n+p}\,\sigma_{n,p}}
  \end{equation*}
  converges in distribution to a Tracy-Widom distribution as
  $n,p\to\infty$ such that
  $n/p\to\gamma\in [0;\infty]$, and
  ${m}/p\to\kappa\in [1;\infty[$. 
\end{theorem}

\noindent{\em Proof } 
The largest eigenvalue of the inverse Wishart MAP estimator is given by
\begin{align*}
  \lambda_{\max}^{(1)}=\frac 1{n+{m}+p+1}
  (n\lambda_{\max}(\boldsymbol S)+\alpha).
\end{align*}
Consequently, 
\begin{align*}
  \frac {\lambda_{\max}^{(1)}-\frac
    n{n+{m}+p+1}\,\mu_{n,p}}
  {\frac{n}{n+p+{m}}\,\sigma_{n,p}}
  &=\frac {\frac
    n{n+{m}+p+1}}{\frac{n}{n+p+{m}}}\frac{\lambda_{\max}(\boldsymbol S)-\mu_{n,p}}{\sigma_{n,p}}
  +\frac{\frac{\alpha}{n+{m}+p+1}}{\frac{n}{n+p+{m}}\sigma_{n,p}},
\end{align*}
converges to a Tracy-Widom distribution, as
$n\sigma_{n,p}\to\infty$.

A more indirect argument is needed for the general case. Recall that
the eigenvalues solves \eqref{eq:qiw.MAP.eigenvalues}, and  
that the derivative \eqref{eq:implicit.diff} is
positive. 
This implies that $\lambda_{\max}^{(q)}$ solves
\eqref{eq:qiw.MAP.eigenvalues} for $\lambda_i(\boldsymbol
S)=\lambda_{\max}(\boldsymbol S)$. Hence,
\begin{equation}\label{eq:defining.eqn}
  (n+p+q{m}+1)\left(\lambda_{\max}^{(q)}
    -\frac n{n+p+q{m}+1}\lambda_{\max}(\boldsymbol S)\right)\left(\lambda_{\max}^{(q)}\right)^{q-1}=\alpha^q q.
\end{equation}
Write 
\begin{multline*}
  \frac n{n+p+q{m}+1}\lambda_{\max}(\boldsymbol S)
  =\frac n{n+p+q{m}+1}\,\sigma_{n,p}\cdot\frac{\lambda_{\max}(\boldsymbol
    S)-\mu_{n,p}}{\sigma_{n,p}}\\
  +\frac n{n+p+q{m}+1}\mu_{n,p},
\end{multline*}
and observe that the first term is $o_P(1)$ whereas the second term
converges to a positive constant.  Thus by the lower bound
\eqref{eq:qiw.eigenval.lower.bound} it follows that
$\lambda_{\max}^{(q)}$ is bounded away from 0 in probability.
Consequently, we obtain
\begin{align*}
  \frac{\lambda_{\max}^{(q)}
    -\frac n{n+p+q{m}+1}\lambda_{\max}(\boldsymbol S)}{\sigma_{n,p}}=O_P\left(\frac{\sigma_{n,p}}{n+p+q{m}+1}\right)=o_P(1)
\end{align*}
from \eqref{eq:defining.eqn}, and hence
\begin{align*}
  \frac{\lambda_{\max}^{(q)}
    -\frac n{n+p+q{m}+1}\mu_{n,p}}{\frac n{n+p+q{m}}\sigma_{n,p}}=\frac{\frac n{n+p+q{m}+1}}{\frac n{n+p+q{m}}}\cdot\frac{\lambda_{\max}(\boldsymbol S)-\mu_{n,p}}{\sigma_{n,p}}+o_P(1)
\end{align*}
converges to a Tracy-Widom distribution.\qed
\\\\
\emph{Remark.} Recall that ${m}\ge p$ so that
${m}$ must increase at least as fast as $p$. Hence in
theorem \ref{thm:tracy-widom}, $\kappa$ cannot be smaller than 1. A
finite value of $\kappa$ means that $q$ increases at the same
rate as $p$ whereas $\kappa=\infty$ would mean that $q$
increases at a faster rate. Note that our result does not include this
scenario.
\\\\
It follows from the proof of theorem \ref{thm:tracy-widom} that
\begin{align*}
  \lambda_{\max}^{(q)}\stackrel{P}{\longrightarrow}\lim_{n,p\to\infty}\frac n{n+p+q{m}+1}\mu_{n,p}
  =1+\frac{2\sqrt{\gamma}-q\kappa}{1+\gamma+q\kappa},
\end{align*}
where the last term is interpreted as 0, if $\gamma$ equals $\infty$.
Thus, the maximal asymptotic bias is smaller than 1. We note that the
asymptotic bias of the largest eigenvalue of the power inverse Wishart
MAP is bounded, whereas the asymptotic bias of the largest eigenvalue
of the MLE is unbounded.  In cases where $p<n$ (so that
$\gamma>1$) we may actually choose $q$ and ${m}$ such that
the asymptotic bias is 0.  Furthermore, the rate of convergence of the
largest eigenvalue of the power inverse Wishart MAP is never slower
than the rate of convergence of the largest eigenvalue of the MLE.
\\\\
\emph{Remark.} It is not obvious how to extend this result to the
case, when $\boldsymbol\Psi$ is not of the form $\alpha\boldsymbol I$, since in this
case the largest eigenvalue of the MAP estimator is not a simple
function of the largest eigenvalue of the MLE. A related question is,
what happens to the asymptotic results, when the covariance matrix of
the underlying normally distributed data is $\boldsymbol\Sigma$ instead of
$\boldsymbol I$. In this case the largest eigenvalue of
$\boldsymbol\Sigma^{-1/2}\boldsymbol S\boldsymbol\Sigma^{-1/2}$ has an asymptotic
Tracy-Widom distribution. As
\begin{align*}
  \lambda_{\max}(\boldsymbol\Sigma^{-1/2}\boldsymbol
  S\boldsymbol\Sigma^{-1/2})\lambda_{\min}(\boldsymbol\Sigma)\le
  \lambda_{\max}(\boldsymbol S)\le \lambda_{\max}(\boldsymbol\Sigma^{-1/2}\boldsymbol
  S\boldsymbol\Sigma^{-1/2})\lambda_{\max}(\boldsymbol\Sigma),
\end{align*}
the asymptotic distribution of $\lambda_{\max}(\boldsymbol S)$ depends on how
the eigenvalues of $\boldsymbol\Sigma$ depends on $p$.

\section{Simulations}\label{sec:simulations}

To investigate the finite sample behaviour of our estimators we report
on a small simulation study. We only consider the MLE, the usual
inverse Wishart MAP and a power inverse Wishart MAP with $q=2$. Both
MAPs are based on priors with $\boldsymbol\Psi=\alpha\boldsymbol I$.

\begin{figure}
  \centering
  \includegraphics[width=0.5\textwidth]{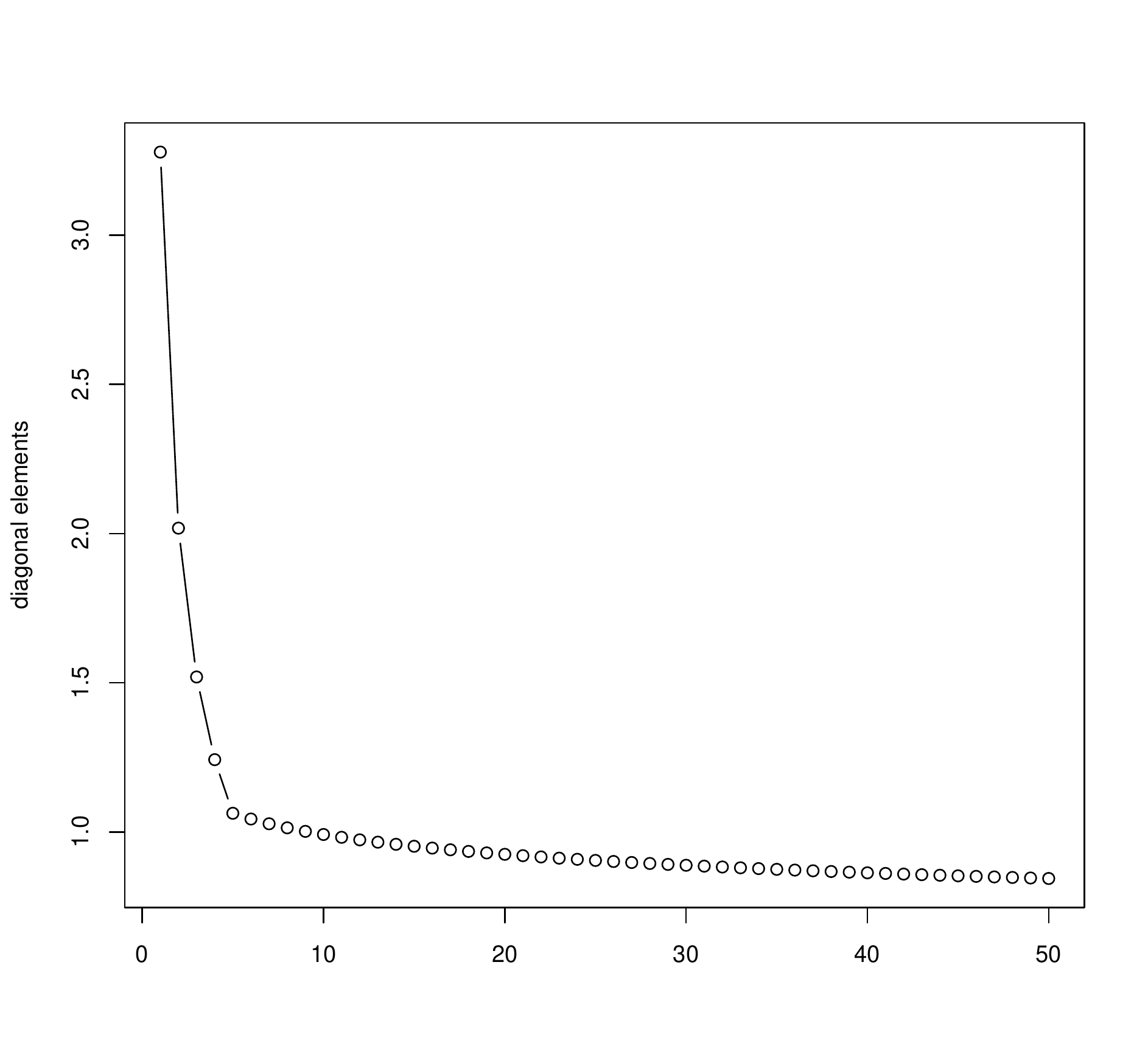}
  \caption{Eigenvalues of $\boldsymbol{\Sigma}$ for $p=50$ used in the simulation.}
  \label{fig:syntheticEigenvalues}
\end{figure}

We consider two types of covariance matrices: The first is
$\boldsymbol{\Sigma}=\boldsymbol{I}$, the second is a diagonal matrix with
diagonal elements equal to
\begin{equation}
  \label{eq:syntheticEigenvalues}
  \boldsymbol{\Sigma}_{ii}=\begin{cases}
    \sigma^2\,i^{-0.7} & \text{for $1\leq  i\le p/10$},\\
    \sigma^2\left(\frac{p}{10}\right)^{-0.6}i^{-0.1} & \text{for $p/10<i\leq p$},
  \end{cases}
\end{equation}
which is illustrated in figure~\ref{fig:syntheticEigenvalues}. Here
there are a few large eigenvalues, but after a steep decrease the
remaining eigenvalues are small and only decrease slowly. This
covariance matrix is chosen to loosely mimic the behaviour of the eigenvalues
in the real data example in the following section.  We consider the
behaviour of the MAP estimators under the quadratic loss function
\begin{align*}
  L_2(\boldsymbol{\Sigma},\boldsymbol{\hat{\Sigma}})
  &=\operatorname{tr}\left({(\boldsymbol{\Sigma}-\boldsymbol{\hat{\Sigma}})
    (\boldsymbol{\Sigma}-\boldsymbol{\hat{\Sigma}})^\top}\right).
\end{align*}
The risk of the MLE and the inverse Wishart MAP can be calculated
explicitly (see \ref{sec:appendix}), but the risk of the power inverse
Wishart MAP cannot, so we rely on simulations.  We will give results
for three choices of $p$, namely 10, 50 and 100. For each value
of $p$, we will use $n=p/2,p,2p$ to
investigate the behaviour in three different ``asymptotic scenarios''.

We first note that it is sufficient to consider diagonal matrices for
$\boldsymbol\Sigma$: For any orthonormal matrix $\boldsymbol V$ we have 
\begin{align*}
  L_2(\boldsymbol{\Sigma},\boldsymbol{\hat{\Sigma}})
  &=L_2(\boldsymbol V^\top\boldsymbol{\Sigma}\boldsymbol V,\boldsymbol V^\top\boldsymbol{\hat{\Sigma}}\boldsymbol
  V),
\end{align*}
and since
\begin{multline*}
  \boldsymbol V^\top\boldsymbol{\hat{\Sigma}}\boldsymbol V=
  \frac n{2({n+p+2{m}+1})}
        \bigg(\boldsymbol V^\top\boldsymbol{S}\boldsymbol V\\
      +\big((\boldsymbol V^\top\boldsymbol{S}\boldsymbol V)^2+8\frac{{n+p+2{m}+1}}{n^2}\alpha\boldsymbol{I}\big)^{1/2}\,\bigg)
\end{multline*}
for the 2-power inverse Wishart MAP (and with a similar result for the
inverse Wishart MAP), the risks are left unchanged by rotations.

In our simulations, we choose $\sigma^2$ in
\eqref{eq:syntheticEigenvalues}, such that the risks of the MLE for
given values of $p$ and $n$ are the same in the two
examples; see table~\ref{tab:MLE.risk} for the values of these risks.
We do not vary the variance parameter $\sigma^2$ in the simulations,
because increasing $\sigma^2$ will give the same results as keeping it
fixed while lowering the floor and scaling the resulting risks. Thus
it is sufficient to vary the floor.
\begin{table}
  \centering
  \begin{tabular}{|l|r@{}lr@{}lr@{}l|}
    \hline&&&&&&\\[-.6em]
    Risk&\multicolumn{2}{c}{$p=10$}&\multicolumn{2}{c}{$p=50$}&\multicolumn{2}{c|}{$p=100$}\\
    \hline&&&&&&\\[-.8em]
    $n=p/2$&18&&98&&198&\\
    $n=p$&10&&50&&100&\\
    $n=2p$&5&.25&25&.25&50&.25\\\hline
  \end{tabular}
  \caption{Quadratic risk of the MLE.}
  \label{tab:MLE.risk}
\end{table}

When comparing the two MAPs, the choice of hyperparameters is crucial:
By choosing suitably different hyperparameters we can easily make one
MAP looks superior to the other. To avoid this we try to choose the
hyperparameters of the inverse Wishart prior so that the two MAP
estimators have the same floor and the same overall amount of
shrinkage.  We believe that a reasonable comparison should use the
same floor. However, if we use the same floor and the same shrinkage
factor, then the regularization curve for the power inverse Wishart MAP
\eqref{eq:SIW.reg.fct} will be below the regularization curve for the
inverse Wishart MAP \eqref{eq:IW.reg.fct} and our simulation results
would be more a consequence of different amounts of shrinking rather
than of the difference between the estimators.  In order to circumvent
this effect, we write \eqref{eq:IW.reg.fct} and \eqref{eq:SIW.reg.fct}
as $a'\lambda+b'$ and $a\lambda+a\sqrt{\lambda^2+b}$ respectively.
Here $a$ and $b$ are functions of the the chosen floor and shrinkage
of the 2-power inverse Wishart MAP, and $b'$ is just the common value
of the floor. For chosen values of floor and shrinkage for the power
inverse Wishart MAP, we choose $a'$, such that
\begin{align*}
  0=&\int_0^L\left(a\lambda+a\sqrt{\lambda^2+b}-a'\lambda-b'\right)d\lambda
\end{align*}
for a suitable value of $L$.  Using $L=\infty$ leads to $a'=2a$, i.e.\
the same shrinkage factor for the MAPs, so we need to choose a finite
value of $L$. We choose $L$ equal to the 99\%-quantile in the
distribution of the largest eigenvalue of the MLE. In this way the two
MAPs has the same ``average regularization'' over the plausible range
of observed eigenvalues.

The shrinkage factors of both MAPs are bounded by the fact that
${m}\ge p$. We use the maximal shrinkage factor for the
power inverse Wishart MAP as well as factors 10\% and 20\% smaller.
 We also use three different values for
the floor --0.8, 1, and 1.2-- corresponding to the average value of
the eigenvalues of $\boldsymbol\Sigma$ (to two decimal places for the matrix
given by \eqref{eq:syntheticEigenvalues}) and values 20\% smaller and larger.

The results based on 5,000 simulations are given in
table~\ref{tab:Identity} and~\ref{tab:Eigenvalues}. The differences
between the two MAPs are small compared to the improvement over the
MLE (see table~\ref{tab:MLE.risk}). This is not unexpected. We have
chosen the hyperparameters of the priors in order to make the MAP
estimators as similar as possible, and all our simulations are in
situations, where the MLE is not expected to work well. We see that
choosing the floor equal to 1 typically leads to smaller risks. This
is not surprising for the $\boldsymbol\Sigma=\boldsymbol I$ case, where all
eigenvalues are equal to 1. Indeed, in this case it is optimal to use
a floor equal to 1 ($\alpha=n+{m}+p+1$) and
shrink as much as possible (${m}\to\infty$). But it is also
the case for the more realistic example, where most of the true
eigenvalues are smaller than 1. Thus, it seems overall beneficial to
overestimate small eigenvalues to some extent. On the other hand, as
one would expect, it is also clear in our simulations that a floor
that is ``too small'' is preferably to one that is ``too large''.

\begin{table}
  \begin{center}
    {\footnotesize \input{table2}}
  \end{center}
  \caption{Quadratic risk, $\boldsymbol{\Sigma}=\boldsymbol{I}$. Lines with $q=1$
    are for an inverse Wishart MAP, $q=2$ for the power inverse
    Wishart MAP. The ``shrink'' is the factor multiplied onto the
    maximally possible shrinkage factor for the power inverse
    Wishart MAP.  The smallest risk for each combination of $(p,
    n)$ is given in bold; the smallest risk for each
    combination of $(p, n)$ and floor and shrinkage is given in italics.}
  \label{tab:Identity}
\end{table}

\begin{table}
  \begin{center}
    {\footnotesize \input{table3}}
  \end{center}
  \caption{Quadratic risk, $\boldsymbol{\Sigma}$ given by
    \eqref{eq:syntheticEigenvalues}. Lines with $q=1$ are for an
    inverse Wishart MAP, $q=2$ for the power inverse Wishart
    MAP. The ``shrink'' is the factor multiplied onto the maximally
    possible shrinkage factor for the power inverse Wishart
    MAP.  The smallest risk for each combination of $(p,
    n)$ is given in bold; the smallest risk for each
    combination of $(p, n)$ and floor and shrinkage is given in italics.}
  \label{tab:Eigenvalues}
\end{table}

For the values used here, more shrinkage (smaller values of the
shrinkage factor) generally leads to smaller risk, regardless of the floor for
the values used here. Obviously, there will be a limit to this effect:
If the floor is too low or too high, too much shrinking will lead to
higher risks due to estimates that are too small or too large.

Overall the power inverse Wishart MAP performs better than the usual
inverse Wishart MAP, when the floor is not too low. It should also be
clear that we cannot conclude that the power inverse Wishart MAP is
always better than the usual inverse Wishart MAP. Along with the other
hyperparameters, the power $q$ must be chosen by the data analyst.

\section{Application to real data}\label{sec:notch.shape}
We consider the data set analysed by \citet{shepstone}, who studied the intercondylar notch in human
osteoarthritic and non-osteoarthritic femora.  The authors considered
96 human femora from a large skeletal population. The femora were annotated
by sex as well as distal eburnation.  The available data is a sampling
of a 2-dimensional spline curve approximation of the silhouette of the
condyle in 50 arch length equidistant points normalised to start in
(0,0) and end in (1,0).

\begin{figure}
  \centering
  \includegraphics[width=.7\textwidth]{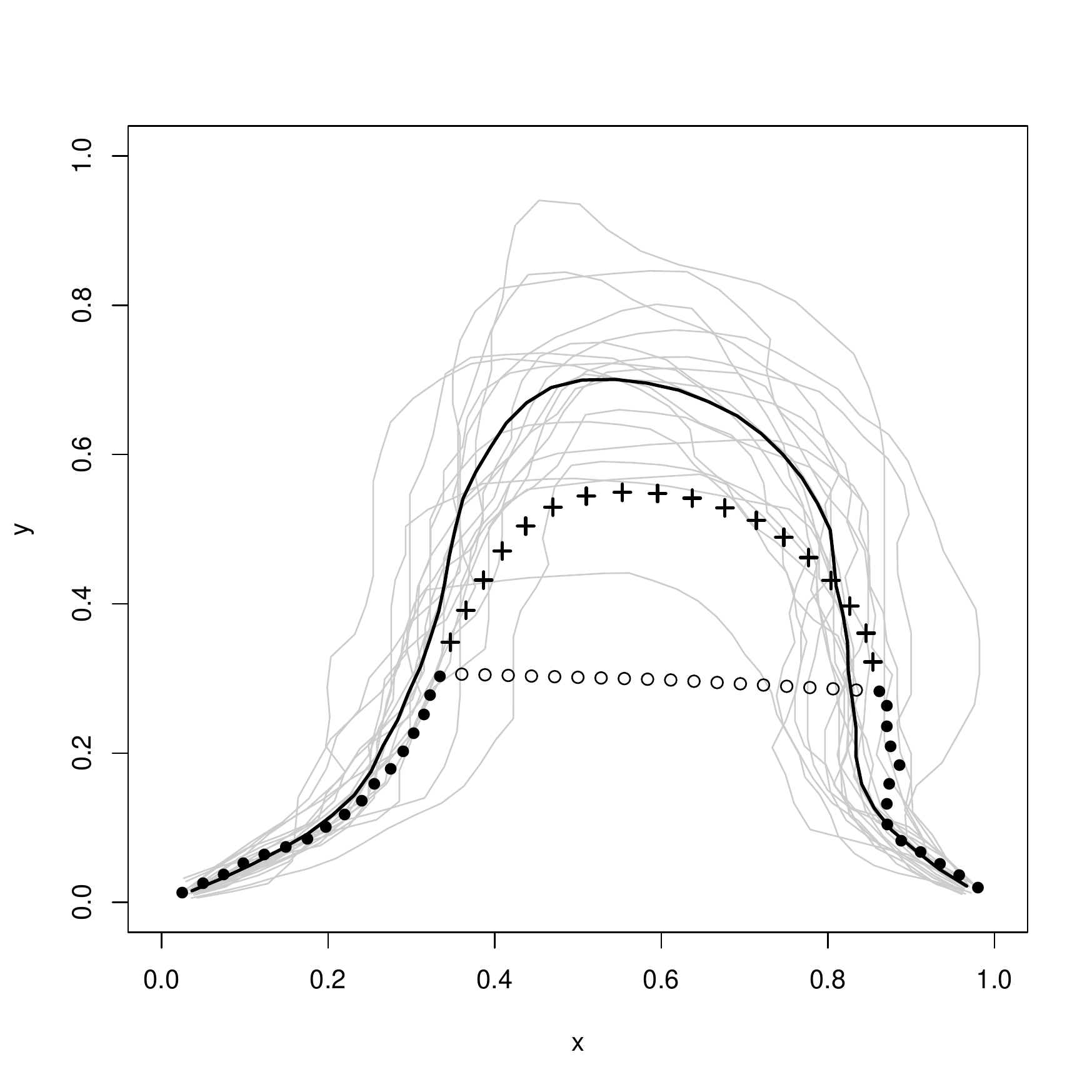}
  \caption{Prediction of the middle part of notch "2283R". The full
    line is the mean shape of the notches, the grey lines the observed
    notches. The circles and bullets outline notch "2283R" as it is in
    the data. We interpret the bullets as observed and the circles as
    missing. The prediction of the missing part is given by the
    pluses.}
  \label{fig:notch.prediction}
\end{figure}

We only consider a part of the data set, namely the 21 condyles with
signs of distal eburnation. One of these (marked "2283R" in the data)
differs markedly from the rest of the condyles (see
figure~\ref{fig:notch.prediction}), and we omit it from the estimation
procedure.  Later we will use the estimated covariance matrix to find
a prediction of this condyle treating the middle part as missing.  In
this application, $n=20$ whereas $p=96$ (two times 50
minus the two end points, which are fixed).

In data like these, it would be natural to expect adjacent $x$ (or
$y$) coordinates to be highly correlated and distant $x$ ($y$)
coordinates to be less correlated, so we will let our choice of
$\boldsymbol\Psi$ reflect this. The $x$ and $y$ coordinates may also be
correlated, but we expect this correlation to be smaller, and we are
not sure of its sign and put this part of the hyperparameter
$\boldsymbol\Psi$ equal to 0.  Also for simplicity, we assume variance
homogeneity in our prior even though it is clear from the fact that
the outlines of the notches have been ``tied down'' at the ends, that there
will be less variation near the ends than in the middle. These
considerations lead to $\boldsymbol\Psi=\alpha\boldsymbol\Psi_0$ with
\begin{equation*}
  \boldsymbol\Psi_0=
  \begin{bmatrix}
    \textbf{AR}(1)_\rho&\boldsymbol{0}\\
    \boldsymbol{0}&\textbf{AR}(1)_\rho
  \end{bmatrix},
\end{equation*}
where \textbf{AR}(1)$_\rho$ is a correlation matrix for an
AR(1)-process with parameter $\rho$, i.e.\ a matrix with $(i,j)$th
element equal to $\rho^{|i-j|}$, and $\boldsymbol{0}$ is a matrix of 0s. Thus,
we use the same correlation parameter for both $x$ and $y$ coordinates
as well as assume variance homogeneity. This may be too simplistic,
but without strong prior beliefs we prefer to keep $\boldsymbol\Psi$ simple.
We use a prior with $q=2$ and ${m}=p$; larger values of
$q$ and ${m}$ leads to smaller shrinkage factors, and with
$p$ considerably larger than $n$ we expect that this
will give a sufficient amount of shrinkage.  The values of
$\rho\,(=0.94)$ and $\alpha\,(=0.02535)$ are chosen by predictive
cross validation \cite{gelfand} using importance sampling.

\begin{figure}
  \centering
  \includegraphics[width=\textwidth]{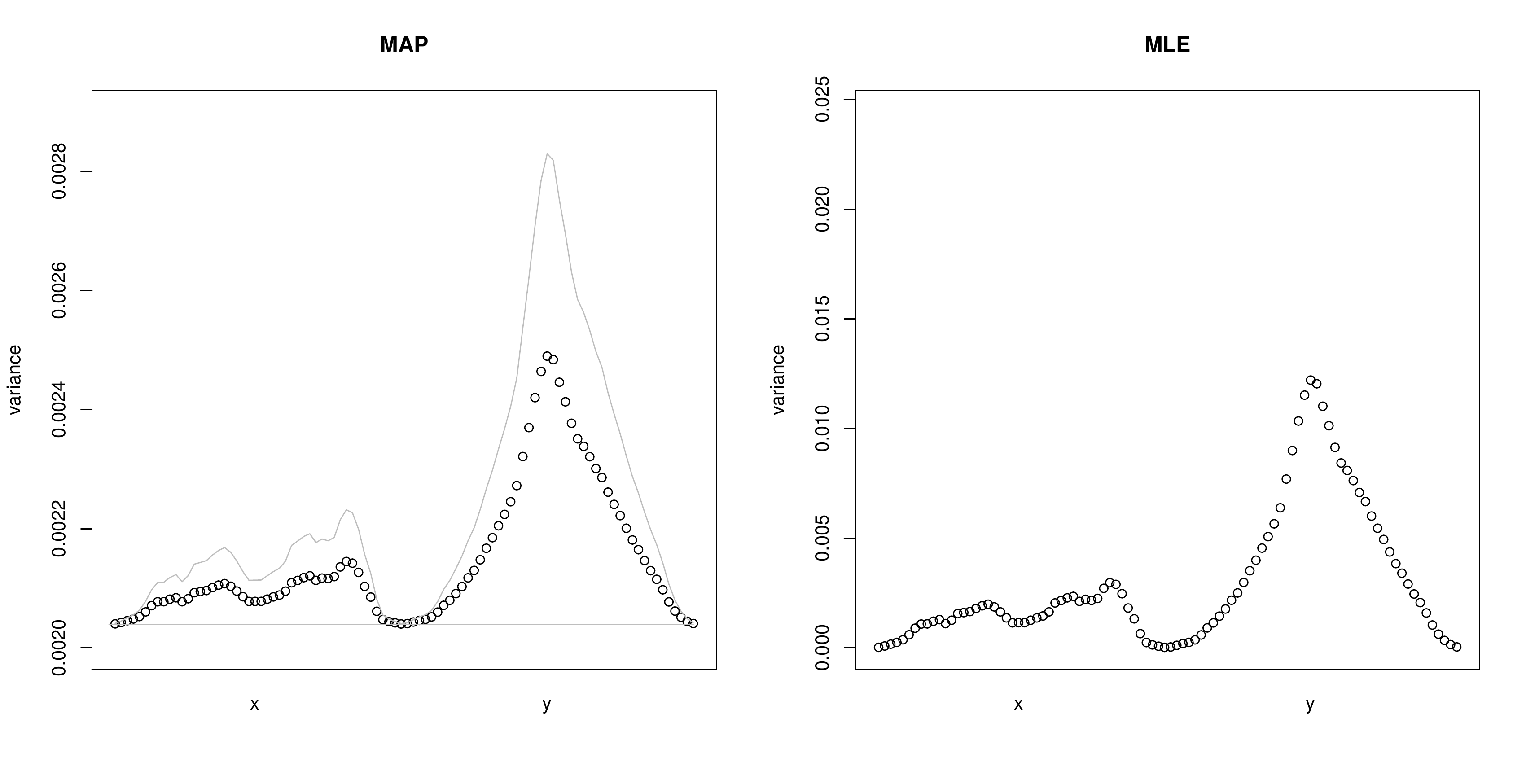}
  \caption{Estimated variances, the MLE on the right, the MAP on the
    left. The grey lines in the plot on the left are the bounds from
    \eqref{eq:piw.bounds}.}
  \label{fig:notch.variances}
\end{figure}

Figure~\ref{fig:notch.variances} shows how the estimated variances are
lifted (by the floor) and shrunken, but also that the relative
relationship between the variances are more or less unchanged. The MAP
estimators of the large variances are much smaller than the MLEs,
which of course is an effect of $\alpha$ and the shrinkage factor
being fairly small; by \eqref{eq:prior.mode} the prior mode is located
at $0.0021{\boldsymbol\Psi_0}$. The smaller variances are lifted, and the
averages of the estimated variances (the traces of the estimators) are
not markedly different (0.0021 for the MAP and 0.0029 for the MLE).

\begin{figure}
  \centering
  \includegraphics[width=\textwidth]{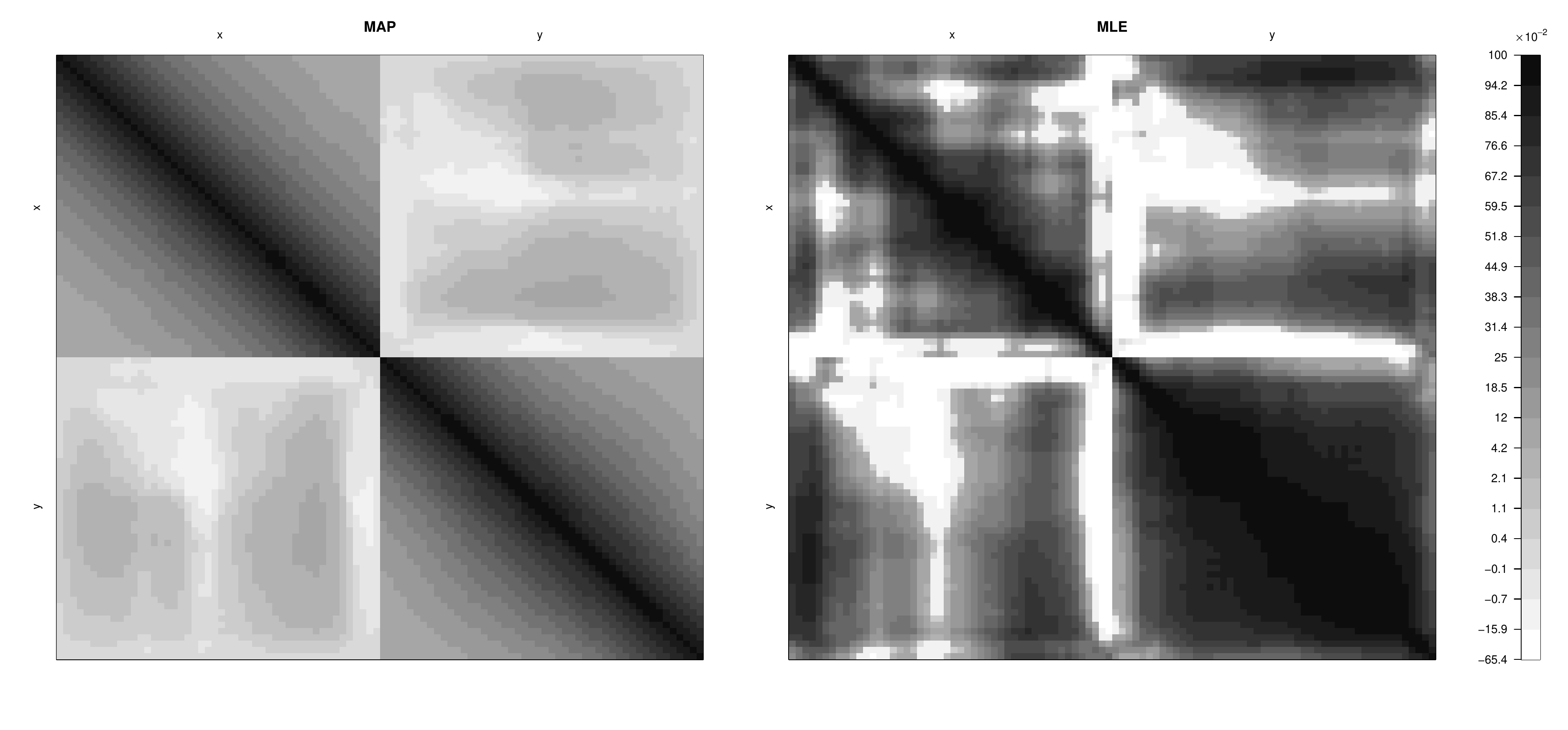}
  \caption{Estimated correlation matrices; MLE on the right, MAP on
    the left. The grey tone-bands are based on the 5\%-, 10\%-,
    \ldots, 95\%-quantiles of the elements of the two estimators.}
  \label{fig:heatmaps.correlations}
\end{figure}

Turning next to the estimated correlation matrix
(figure~\ref{fig:heatmaps.correlations}), we see how the prior
independence of $x$ and $y$ coordinates removes most of the
correlation between $x$ and $y$ coordinates. The prior's
AR(1)-structure is also evident in the correlations between $x$
coordinates and between the $y$ coordinates.

\begin{figure}
  \centering
  \includegraphics[width=\textwidth]{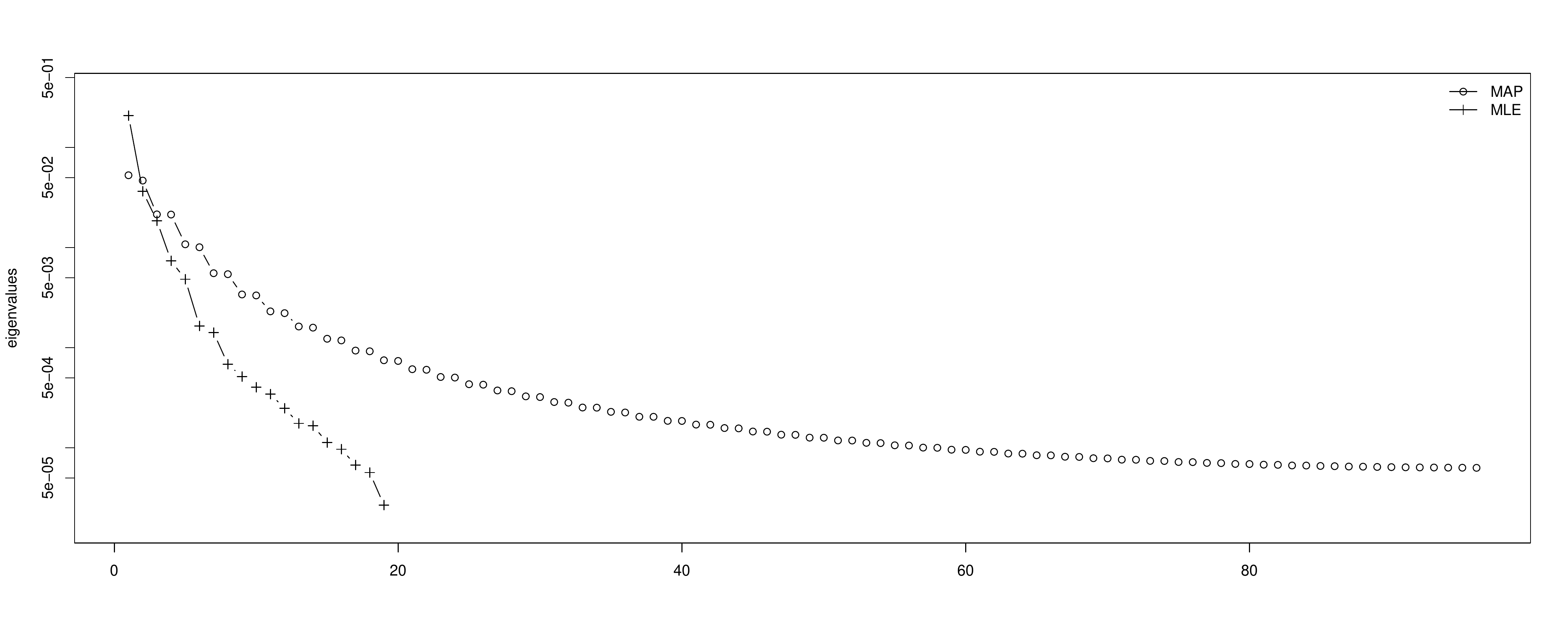}
  \caption{Eigenvalues of the MLE ('$\scriptscriptstyle +$') and the
    MAP ('$\scriptscriptstyle\circ$'); note that the y-axis is
    logarithmic.}
  \label{fig:notch.eigenvals}
\end{figure}

\begin{figure}
  \centering
  \includegraphics[width=\textwidth]{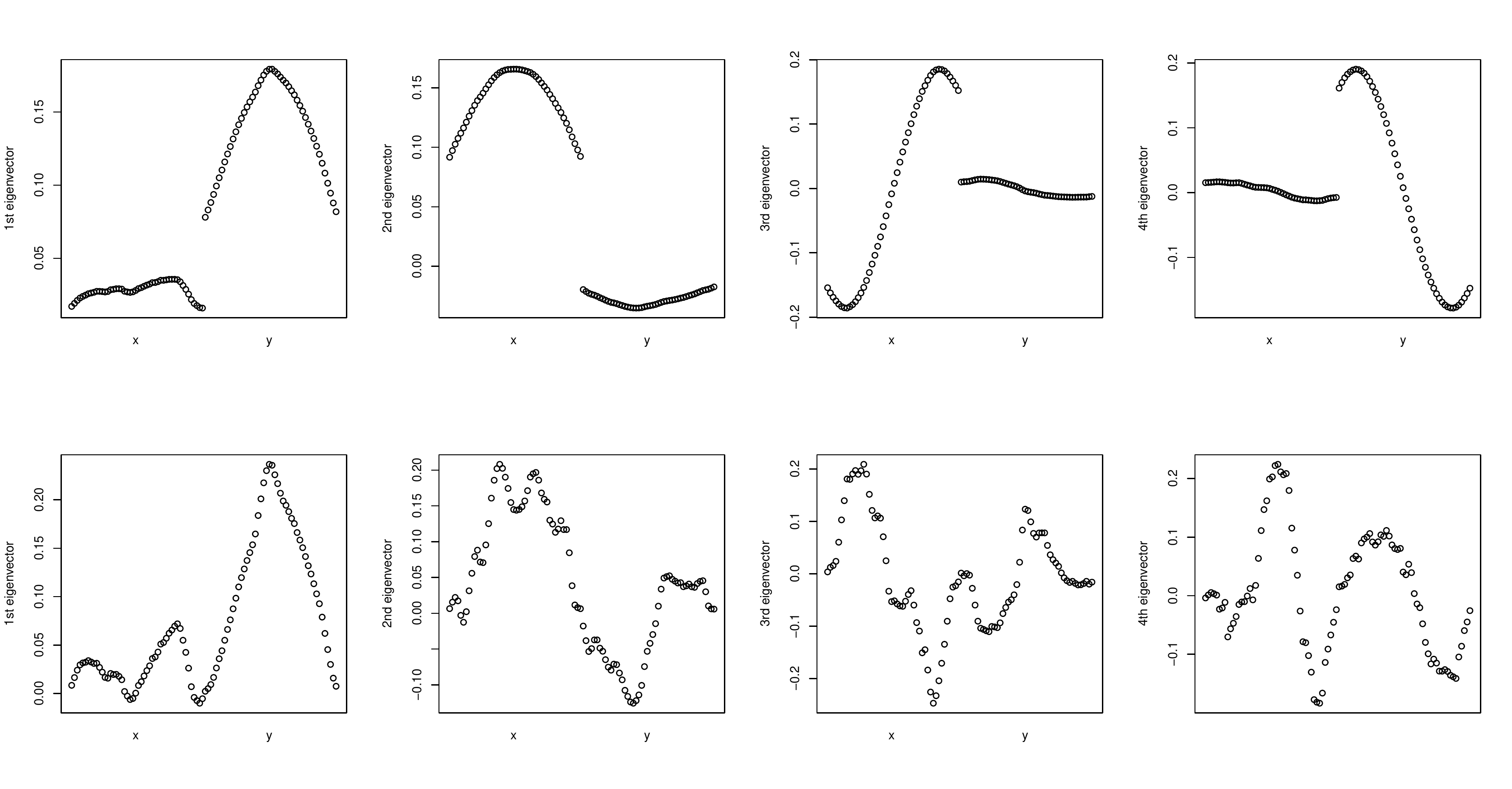}
  \caption{First 4 eigenfunctions of the estimators. The top row is
    the MAP, the bottom row is the MLE.}
  \label{fig:notch.eigenfcts}
\end{figure}

The eigenvalues and the first four eigenvectors of the MAP and the MLE
are shown in figures~\ref{fig:notch.eigenvals}
and~\ref{fig:notch.eigenfcts}. We see that the prior lifts the
eigenvalues; only the largest eigenvalue is smaller when estimated by
the MAP, than when it is estimated by the MLE. Note that the $y$-axis
in figure~\ref{fig:notch.eigenvals} is logarithmic, so that the
difference between the largest eigenvalues of the two estimators is
rather big.  The eigenvalues of the MAP estimator are pairwise
similar. This is probably an effect of the block-diagonal $\boldsymbol\Psi$;
it tends to split the variation into a part mostly related to the
$x$-coordinates and a part mostly related to the $y$-coordinates. This
is also what we see from figure~\ref{fig:notch.eigenfcts}. Indeed it
seems that the sinusoidal-looking eigenvectors of AR(1)-correlation
matrices and the block diagonal form have a dominant effect on the
resulting MAP estimator.

Any application of MAP estimation is a compromise between the data and
the prior: We wish to balance the information provided by the data
with the stability introduced by the prior.  It is not surprising that
the prior has a large effect in this example. Even if we suspect that
the true covariance matrix is more complicated, there hardly is any
information in the data to help us discover it. The size of dataset is
very small compared to the dimension of the unknown covariance matrix,
so the shrinkage factor is quite small, and $\boldsymbol\Psi$ has a lot of
weight in the resulting estimator.  Though this is the intended effect
of MAP estimation, it also means that the prior should be chosen
carefully. In this example we have used a very simple choice of
$\boldsymbol\Psi$. More complicated choices may be considered: Different
variances for $x$- and $y$-coordinates, as well as correlation between
$x$ and $y$-coordinates are easily implemented in the estimator.
However, choosing the values of the hyperparameters is more
complicated. Our solution to this problem is basically a grid search,
and the more parameters that need to be chosen, the longer the
computation time.  For this reason, we will not attempt a more
complicated prior for this example.

As mentioned at the beginning of this section the condyle "2283R"
differs radically from the rest. As seen in figure~\ref{fig:notch.prediction}, where it is represented by circles and
bullets, it seems to have had its middle part ''cut off'', when compared
to the other condyles in the dataset (grey curves in the figure). As an
illustration we pretend that the middle part (the part of the condyle
represented by the circles) are missing data and try to predict it.
The usual EBLUP formula (see e.g.\ \citet[p.\ 37]{anderson03}) based on
the MLE breaks down; there are 30 observed points (the bullets in the
figure), so with only 20 fully observed condyles the covariance matrix
corresponding to the observed part of "2283R" is singular and cannot
be inverted.  The MAP, on the other hand, is regular, and when using
this in the formula, we obtain the prediction given in figure~\ref{fig:notch.prediction} by the pluses.

\section{Conclusion}

In this paper we have introduced a new class of distributions --the
power inverse Wishart distributions-- on the set of positive definite
matrices. Used as priors for unknown covariance matrices of
multivariate Gaussian data, they lead to easily calculable maximum a
posteriori estimators. Our simulations suggest that the MAP estimators
perform better than the MLE in terms of overall quadratic risk. We
have derived some asymptotic properties of these estimators and have
seen that these are as good as or in some situations even better than
those of the MLE.

As we have seen in sections \ref{sec:simulations} and \ref{sec:notch.shape}
the choice of prior influences the MAP estimator. Obviously, if this
was not the case, there would be little reason for using the MAP
estimator. On the other hand, it also means that the prior should be
chosen carefully. In section \ref{sec:notch.shape} we chose the form
of the prior mode based on prior beliefs but the values of it was
determined by cross validation. Our implementation of this cross
validation is too computationally demanding to allow a further
investigation of its properties, so it is difficult to know if this is
in any sense optimal. Clearly, this is an area that requires
additional work. 

It is quite easy to extend our results (except theorem
\ref{thm:density}) to improper priors with ${m}<p$; in
theorem \ref{thm:tracy-widom} this would allow $\kappa$ to be any
non-negative real. By allowing improper priors, we could obtain a MAP
estimator in our example with less shrinkage than the one we have used.
It is less obvious whether our results can be extended to
values of $q$ that are not positive integers, as many of our arguments
rely on $q$ being a positive integer. 

We hope that the additional flexibility provided by the power inverse Wishart
MAP will prove to be useful when estimating large covariance
matrices based on limited amounts of data. 

\section*{Acknowledgments}
We gratefully acknowledge the funding from the Danish Research
Foundation (Den Danske Forskningsfond) supporting this work, and Bo
Markussen, University of Copenhagen, for useful comments.

\appendix
\section{Quadratic risk}
\label{sec:appendix}
 
For estimators of the form
\begin{equation*}
  \hat{\boldsymbol{\Sigma}}=a\boldsymbol{S}+b\boldsymbol{I}
\end{equation*}
such as the MLE and the inverse Wishart MAP estimator, the expected
quadratic risk is
\begin{align*}
   E[L_2(\boldsymbol{\Sigma},\boldsymbol{\hat{\Sigma}})]
  &=\operatorname{tr}\left( {E\left[\left(a\boldsymbol{S}+b\boldsymbol{I}-\boldsymbol\Sigma\right)^2\right]}\right)\\
  &=a^2\left(\frac{n-1}{n}\sum_{i=1}^p\sum_{j=1}^p 
   \Sigma_{i,j}^2
   +\frac{n-1}{n^2}\left(\sum_{i=1}^p\Sigma_{i,i}\right)^2\right)\\
  &\qquad+2a\frac{n-1}{n}\left(b\,\operatorname{tr}\left({\boldsymbol\Sigma}-\operatorname{tr}\left({\boldsymbol\Sigma^2}\right)\right)\right)\\
  &\qquad+b^2p-2b\operatorname{tr}\left({\boldsymbol\Sigma}\right)+\operatorname{tr}\left({\boldsymbol\Sigma^2}\right)
\end{align*}
as $S$ is Wishart distributed with parameters $(n-1,{\boldsymbol\Sigma}/n)$. 

The expression for
$E[L_2(\boldsymbol{\Sigma},\boldsymbol{\hat{\Sigma}})]$ is a convex polynomial of
$(a,b)$ of degree two and thus has a minimal value. Thus, there are
unique optimal values of the floor and the shrinkage for the inverse
Wishart MAP (for a given $\boldsymbol\Sigma$), but there are also choices
that will lead to inverse Wishart MAPs with larger risks than the MLE.

\bibliographystyle{elsarticle-harv}
\bibliography{references}

\end{document}

%% file: table2.tex
\begin{tabular}{|cc|rrr|rrr|rrr|}
   \hline   &\multicolumn{1}{c}{floor}&\multicolumn{3}{c}{0.8}&\multicolumn{3}{c}{1}&\multicolumn{3}{c|}{1.2}\\
$(p,n)$&
\hspace{-1em}shrink & 1 & 0.9 & 0.8 & 1 & 0.9 & 0.8 & 1 & 0.9 & 0.8 \\ 
   \hline
(10,5)&q=1 & \textit{0.34} & \textit{0.31} & \textit{0.30} & 0.26 & 0.20 & 0.15 & 0.97 & 0.87 & 0.79 \\ 
  &q=2 & 0.62 & 0.60 & 0.57 & \textit{0.10} & \textit{0.08} & \textbf{{0.07}} & \textit{0.21} & \textit{0.22} & \textit{0.23} \\ 
   \hline
(10,10)&q=1 & \textit{0.40} & \textit{0.33} & \textit{0.27} & 0.67 & 0.52 & 0.39 & 1.70 & 1.48 & 1.29 \\ 
  &q=2 & 0.44 & 0.43 & 0.41 & \textit{0.13} & \textit{0.11} & \textbf{{0.09}} & \textit{0.47} & \textit{0.45} & \textit{0.43} \\ 
   \hline
(10,20)&q=1 & 0.72 & 0.53 & 0.37 & 1.51 & 1.17 & 0.90 & 3.01 & 2.56 & 2.16 \\ 
  &q=2 & \textit{0.30} & \textit{0.27} & \textit{0.25} & \textit{0.32} & \textit{0.26} & \textbf{{0.20}} & \textit{1.00} & \textit{0.91} & \textit{0.84} \\ 
   \hline\hline
(50,25)&q=1 & \textit{1.40} & \textit{1.37} & \textit{1.36} & 0.95 & 0.75 & 0.58 & 4.44 & 4.09 & 3.76 \\ 
  &q=2 & 3.12 & 2.99 & 2.86 & \textit{0.55} & \textit{0.46} & \textbf{{0.37}} & \textit{1.07} & \textit{1.09} & \textit{1.13} \\ 
   \hline
(50,50)&q=1 & \textit{1.27} & \textit{1.13} & \textit{1.05} & 2.08 & 1.64 & 1.26 & 6.77 & 6.05 & 5.39 \\ 
  &q=2 & 2.19 & 2.11 & 2.05 & \textit{0.68} & \textit{0.56} & \textbf{{0.45}} & \textit{2.32} & \textit{2.23} & \textit{2.15} \\ 
   \hline
(50,100)&q=1 & 1.63 & \textit{1.22} & \textit{0.92} & 4.22 & 3.31 & 2.54 & 10.60 & 9.24 & 8.00 \\ 
  &q=2 & \textit{1.42} & 1.31 & 1.23 & \textit{1.56} & \textit{1.26} & \textbf{{1.00}} & \textit{4.93} & \textit{4.53} & \textit{4.14} \\ 
   \hline\hline
(100,50)&q=1 & \textit{2.78} & \textit{2.71} & \textit{2.69} & 1.95 & 1.54 & 1.19 & 9.00 & 8.27 & 7.60 \\ 
  &q=2 & 6.23 & 5.97 & 5.72 & \textit{1.10} & \textit{0.92} & \textbf{{0.75}} & \textit{2.15} & \textit{2.19} & \textit{2.26} \\ 
   \hline
(100,100)&q=1 & \textit{2.52} & \textit{2.24} & \textit{2.07} & 4.20 & 3.31 & 2.54 & 13.65 & 12.18 & 10.85 \\ 
  &q=2 & 4.38 & 4.22 & 4.09 & \textit{1.36} & \textit{1.12} & \textbf{{0.89}} & \textit{4.63} & \textit{4.45} & \textit{4.29} \\ 
   \hline
(100,200)&q=1 & 3.26 & \textit{2.42} & \textit{1.83} & 8.45 & 6.64 & 5.09 & 21.26 & 18.54 & 16.06 \\ 
  &q=2 & \textit{2.83} & 2.60 & 2.45 & \textit{3.11} &
  \textit{2.53} & \textbf{{2.00}} & \textit{9.85} &
  \textit{9.05} & \textit{8.28} \\ 
   \hline
\end{tabular}

%% file: table3.tex
\begin{tabular}{|cc|rrr|rrr|rrr|}
   \hline   &\multicolumn{1}{c}{floor}&\multicolumn{3}{c}{0.8}&\multicolumn{3}{c}{1}&\multicolumn{3}{c|}{1.2}\\
$(p,n)$&
\hspace{-1em}shrink & 1 & 0.9 & 0.8 & 1 & 0.9 & 0.8 & 1 & 0.9 & 0.8 \\ 
   \hline
(10,5)&q=1 & \textit{0.38} & \textit{0.36} & \textit{0.34} & 0.31 & 0.25 & 0.20 & 1.02 & 0.92 & 0.84 \\ 
  &q=2 & 0.67 & 0.64 & 0.62 & \textit{0.15} & \textit{0.13} & \textbf{{0.12}} & \textit{0.26} & \textit{0.27} & \textit{0.27} \\ 
   \hline
(10,10)&q=1 & \textit{0.44} & \textit{0.36} & \textit{0.31} & 0.70 & 0.56 & 0.44 & 1.73 & 1.51 & 1.33 \\ 
  &q=2 & 0.48 & 0.47 & 0.46 & \textit{0.18} & \textit{0.15} & \textbf{{0.13}} & \textit{0.51} & \textit{0.49} & \textit{0.48} \\ 
   \hline
(10,20)&q=1 & 0.74 & 0.55 & 0.41 & 1.52 & 1.20 & 0.93 & 3.04 & 2.59 & 2.20 \\ 
  &q=2 & \textit{0.33} & \textit{0.30} & \textit{0.29} & \textit{0.35} & \textit{0.29} & \textbf{{0.24}} & \textit{1.04} & \textit{0.95} & \textit{0.88} \\ 
   \hline\hline
(50,25)&q=1 & \textit{7.33} & \textit{7.42} & \textit{7.53} & 6.95 & 6.86 & 6.79 & 10.49 & 10.24 & 10.02 \\ 
  &q=2 & 9.24 & 9.19 & 9.15 & \textit{6.73} & \textit{6.72} & \textbf{{6.70}} & \textit{7.31} & \textit{7.40} & \textit{7.51} \\ 
   \hline
(50,50)&q=1 & \textit{6.38} & \textit{6.44} & \textit{6.56} & 7.28 & 7.02 & 6.84 & 12.02 & 11.50 & 11.02 \\ 
  &q=2 & 7.53 & 7.60 & 7.69 & \textbf{{6.12}} & \textit{6.14} & \textit{6.18} & \textit{7.83} & \textit{7.89} & \textit{7.95} \\ 
   \hline
(50,100)&q=1 & 5.68 & \textit{5.55} & \textit{5.56} & 8.36 & 7.74 & 7.27 & 14.83 & 13.74 & 12.80 \\ 
  &q=2 & \textit{5.59} & 5.72 & 5.91 & \textit{5.88} & \textbf{{5.82}} & \textbf{{5.82}} & \textit{9.38} & \textit{9.20} & \textit{9.06} \\ 
   \hline\hline
(100,50)&q=1 & \textit{25.71} & \textit{26.11} & \textit{26.55} & 25.06 & 25.09 & 25.18 & 32.24 & 31.94 & 31.71 \\ 
  &q=2 & 29.81 & 29.89 & 29.98 & \textbf{{24.89}} & \textit{25.01} & \textit{25.16} & \textit{26.08} & \textit{26.42} & \textit{26.81} \\ 
   \hline
(100,100)&q=1 & \textit{22.25} & \textit{22.76} & \textit{23.39} & 24.22 & 24.08 & 24.04 & 33.86 & 33.12 & 32.51 \\ 
  &q=2 & 24.72 & 25.19 & 25.72 & \textbf{{22.09}} & \textit{22.44} & \textit{22.80} & \textit{25.64} & \textit{26.01} & \textit{26.43} \\ 
   \hline
(100,200)&q=1 & 18.81 & 19.15 & \textit{19.73} & 24.39 & 23.71 & 23.31 & 37.48 & 35.84 & 34.49 \\ 
  &q=2 & \textit{18.32} & \textbf{{19.14}} & 20.04 & \textit{19.21} & \textit{19.61} & \textit{20.14} & \textit{26.43} & \textit{26.57} & \textit{26.82} \\ 
   \hline
\end{tabular}